\documentclass[a4paper,11pt]{article}

\usepackage{jcappub} 

\newcommand{\simgt}{\lower.5ex\hbox{$\; \buildrel > \over \sim \;$}}
\newcommand{\simlt}{\lower.5ex\hbox{$\; \buildrel < \over \sim \;$}}

\title{Principal Component Analysis of Modified Gravity using Weak Lensing and Peculiar Velocity Measurements}

\author[a]{Shinsuke Asaba,}
\author[b]{Chiaki Hikage,}
\author[c]{Kazuya Koyama,}
\author[d,c]{Gong-Bo~Zhao,}
\author[e]{Alireza Hojjati}
\author[f,g,c]{and Levon Pogosian}

\affiliation[a]{Department of Physics, Graduate School of Science, Nagoya University,\\ Aichi 464-8602, Japan}
\affiliation[b]{Kobayashi Maskawa Institute (KMI), Nagoya University,\\ Aichi 464-8602, Japan}
\affiliation[c]{Institute of Cosmology and Gravitation, University of Portsmouth,\\ Portsmouth, PO1 3FX, UK}
\affiliation[d]{National Astronomy Observatories, Chinese Academy of Science,\\ Beijing, 100012, P.R.China}
\affiliation[e]{Institute for the Early Universe, Ewha Womans University,\\ Seoul, 120-750, Korea}
\affiliation[f]{Department of Physics, Simon Fraser University, Burnaby,\\ British Columbia, V5A 1S6, Canada}
\affiliation[g]{Centre for Theoretical Cosmology, DAMTP, University of Cambridge,\\ CB3 0WA, UK}

\emailAdd{asaba.shinsuke@j.mbox.nagoya-u.ac.jp}
\emailAdd{hikage@kmi.nagoya-u.ac.jp}
\emailAdd{Kazuya.Koyama@port.ac.uk}
\emailAdd{gong-bo.zhao@port.ac.uk}
\emailAdd{aha25@sfu.ca}
\emailAdd{levon@sfu.ca}

\abstract{We perform a principal component analysis to assess ability of future
observations to measure departures from General Relativity in
predictions of the Poisson and anisotropy equations on linear
scales. In particular, we focus on how the measurements of
redshift-space distortions (RSD) observed from spectroscopic galaxy
redshift surveys will improve the constraints when combined with
lensing tomographic surveys. Assuming a Euclid-like galaxy imaging and
redshift survey, we find that adding the 3D information decreases the
statistical uncertainty by a factor between 3 and 10 compared to the
case when only observables from lensing tomographic surveys are
used. We also find that the number of well-constrained modes increases
by a factor between 3 and 7. Our study indicates the importance of
joint galaxy imaging and redshift surveys such as SuMIRe and Euclid to
give more stringent tests of the $\Lambda$CDM model and to distinguish
between various modified gravity and dark energy models.
}


\arxivnumber{1306.2546}

\begin{document}
\maketitle
\flushbottom

\section{Introduction}\label{sec:1}
Finding the origin of the accelerated cosmic expansion discovered by the distance measurements using Type Ia supernovae (SNe) \cite{A.G.Riessetal:98,S.Perlmutteretal:99} is a major goal in modern cosmology. The current standard model, $\Lambda$ Cold Dark Matter ($\Lambda$CDM), in which dark matter and dark energy (DE) comprise 96\% of the total energy in the universe, can explain the cosmic acceleration and is also supported by other observations, such as cosmic microwave background (CMB) \cite{PLANCK} and galaxy distribution \cite{Sanchez:2012sg}. Various DE models such as quintessence \cite{Wetterich:88}, and K-essence \cite{Chibaetal:00}, have been proposed; however, we do not have direct evidence of the existence of DE. An alternative way to explain the accelerating expansion is modified gravity (MG) in which general relativity (GR) is modified on cosmological scales, as in the tensor-vector-scalar theory of gravity\cite{Bekenstein:04}, Dvali-Gabadadze-Porrati model \cite{DGP}, MOND\cite{Milgrom:83}, Einstein-Aether theory \cite{Elingetal:04}, $f(R)$ model \cite{Capozziello:2003tk,Carrolletal:04}, or Galileon gravity \cite{Nicolisetal:09} (see \cite{Clifton:2011} for a review).

A promising way to distinguish DE from MG models is observing the galaxy distribution and weak lensing in detail in order to track the evolution of matter density fluctuation and the perturbations associated with the metric. Perturbative approaches and numerical simulations have been used to study evolution of cosmological perturbations in MG \cite{Oyaizuetal:08,Zhaoetal:11,Li:2011vk,Puchweinetal:13}.  One can constrain the properties of dark energy and various MG models by comparing predictions of theoretical models with various ongoing and planned galaxy redshift and lensing surveys such as Dark Energy Survey (DES) \cite{DES}, Baryon Oscillation Spectroscopic Survey (BOSS) \cite{BOSS2}, Large Synoptic Survey Telescope (LSST) \cite{LSST}, Subaru Measurement of Images and Redshift (SuMIRe) \cite{SUMIRE}, BigBOSS \cite{BigBOSS,BigBOSS2}, Hobby-Eberly Telescope Dark Energy Experiment (HETDEX) \cite{HETDEX}, and Euclid \cite{euclid2}.

One approach to studying MG is to constrain parameters describing each MG model from observations. However, such model-dependent methods only constrain a finite range of possibilities, and cannot anticipate all types of deviations from GR.
Another approach is to constrain departures from GR in a model-independent way, i.e. to parametrize the Poisson and anisotropy equations describing the relation between metric perturbations and the stress-energy tensor by two functions $\mu$ and $\gamma$, respectively, that depend on $k$ and $z$ \cite{G.B.Zhaoetal:09,Pogosianetal:10} (this parametrization is equivalent to $\tilde{G}_{\rm eff}$-$\eta$ in \cite{Zhangetal:07} and $Q$-$\eta$ in \cite{Amendolaetal:08}; see refs.~\cite{Hu:2007pj, Baker:2011jy, Zuntz:2011aq, Baker:2012zs,Battye:2013er} for alternative approaches). Such a parametrization can be applied to a broad class of MG models, including the $f(R)$ and DGP models.  (Unclustered) DE scenarios based on GR, including the $\Lambda$CDM model, satisfy the condition of $\mu=\gamma=1$ and thereby any significant detection of the departure from unity would falsify $\Lambda$CDM and a broad class of DE models.  It is difficult, however, to constrain two arbitrary functions of two variables, $\mu(k,z)$ and $\gamma(k,z)$, although there were several works in which they were partially constrained by current observations after assuming some functional forms \cite{Zhaoetal:10,Samushiaetal:13,Simpsonetal:13}. Principal Component Analysis (PCA) provides an efficient way to compress the parameter space and forecast the well-constrained independent modes for different types of observations \cite{Zhao:2009fn,Hojjatietal:12,Hall:2012wd}.  A broad class of MG models can be described by some linear combinations of
the eigenmodes, and thus the uncertainties in the eigenmodes can be translated into forecasted constraints on parameters in specific MG models.

We perform a PCA of $\mu(k,z)$ and $\gamma(k,z)$ for lensing
tomographic surveys combined with galaxy redshift surveys.  Several
joint galaxy imaging and redshift surveys, such as SuMIRe
and Euclid, are planned. Here we perform a Fisher analysis to compute
the eigenmodes of $\mu$ and $\gamma$ in ($k,z$) space assuming an
Euclid-like survey. We study features of the principal component modes and
forecast their associated errors.  In
\cite{Zhao:2009fn,Hojjatietal:12}, a PCA was performed for the
upcoming lensing surveys, such as DES and LSST, combined with Planck
and Type Ia SNe dataset.  In this paper, we demonstrate quantitatively and
qualitatively the improvement by adding the spectroscopic galaxy
redshift survey.

This paper is organized as follows. In section~\ref{sec:2}, we introduce the MG parameters and the PCA. In section~\ref{sec:3}, we explain the observables derived from lensing and redshift-space galaxy clustering and describe the experiments assumed in our forecasts and how to perform the Fisher analysis.  In section~\ref{sec:4}, we show the results of the PCA, focusing specially on how the information of the spectroscopic galaxy redshift survey improves the constraints on the parameters. We discuss our results in section~\ref{sec:5}. Section~\ref{sec:6} is devoted to summary and conclusions.

\section{Formalism}\label{sec:2}
\subsection{Parameterization of Modified Gravity}
We study the linear evolution of the matter density fluctuation and metric perturbations around the flat Friedmann-Robertson-Walker metric in
the Newtonian gauge. The line element is given by
\begin{align}
  ds^2=a(\tau)^2[-(1+2\Psi)d\tau^2+(1-2\Phi)d{\bf x}^2], \label{eq:1}
\end{align}
where $\Psi$ and $\Phi$ are, respectively, the gravitational potential and curvature perturbation, and $\tau$ is the
conformal time.  In Fourier space, the linearly perturbed energy-momentum conservation equations for matter are given by
\begin{align}
	\dot{\delta}+\theta-3\dot{\Phi}&=0, \label{eq:2}\\
	\dot{\theta}+\mathcal{H}\theta-k^2\Psi&=0, \label{eq:3}
\end{align}
where $\delta\equiv\delta \rho/\rho$ is the matter density contrast, $\theta\equiv ik^av_a$ is the divergence of the velocity field, the dot indicates differentiation with respect to conformal time $\tau$, and $\mathcal{H}\equiv a^{-1}da/d\tau$.  We need two additional equations to solve for the behavior of the four perturbation variables: $\delta$, $\theta$, $\Psi$, and $\Phi$.  In GR, Einstein's equations set the relation between the matter density and the gravitational potential:
\begin{align}
	k^2\Psi=-4\pi a^2G \rho\Delta, \label{eq:5}
\end{align}
where $\Delta$ is the comoving density perturbation, $\Delta = \delta + 3 \mathcal{H} \theta/k^2$, and the relation between the two metric perturbations:
\begin{align}
	k^2(\Phi-\Psi)=12\pi a^2G(\rho+P)\sigma, \label{eq:6}
\end{align}
where $\sigma$ is the anisotropic stress. In the $\Lambda$CDM model, the anisotropic stress of the matter is negligible during matter
dominated era, and thus $\Phi=\Psi$.  In MG, the relations among matter density and the two metric perturbations can be different from
eqs.~(\ref{eq:5}) and (\ref{eq:6}).  We characterize MG by modifying these relations as follows:
\begin{align}
	k^2\Psi&=-4\pi a^2G \mu(k,z)\rho \Delta, \label{eq:7}\\
	\frac{\Phi}{\Psi}&=\gamma(k,z), \label{eq:8}
\end{align}
where $\mu$ and $\gamma$ are unity for all $k$ and $z$ in GR. So, deviations of $(\mu,\gamma)$ from unity would indicate a deviation from $\Lambda$CDM. From eqs.~(\ref{eq:2}), (\ref{eq:3}) and (\ref{eq:7}), we get the linear evolution equation of the matter density contrast on sub-horizon scales
\begin{align}
	\ddot{\delta}+\mathcal{H}\dot{\delta}-4\pi a^2G\mu(k,z)\rho\delta=0, \label{eq:4}
\end{align}
where the contribution of $\dot{\Phi}$ is small enough to ignore when $\mathcal{H}/k\ll 1$.  In GR ($\mu=1$), the growing mode solution of eq.~(\ref{eq:4}) in the matter dominated era is given by
\begin{align}
	D\propto \frac{\mathcal{H}}{a}\int^a_0\frac{da}{\mathcal{H}^3}, \label{eq:10}
\end{align}
where $D$ is called the linear growth factor and is independent of scale.  In MG, $D$ generally has a different dependence of $k$ and
$z$.  It follows from our parametrization that the relation between the lensing potential $\Phi+\Psi$ and the matter density is given by
\begin{align}
	k^2(\Phi+\Psi)=-8\pi a^2G\Sigma(k,z)\rho\Delta, \label{eq:9}
\end{align}
where $\Sigma=\mu(1+\gamma)/2$.

Lensing tomography is sensitive to the change of lensing potential, which is proportional to $\Sigma$. On the other hand, large-scale anisotropy of galaxy clustering due to the bulk motion of galaxies included in the observed galaxy power spectra in the redshift-space, i.e., redshift-space distortions (RSD) \cite{Kaiser:87} , provides a powerful tool to constrain $\mu$.  Hence, combining RSD measurements with the lensing tomography can reduce the degeneracy between $\mu$ and $\gamma$ \cite{Guzik:2009cm, Songetal:11}.

\subsection{Principal Component Analysis}
The parameter space of $\mu(k,z)$ and $\gamma(k,z)$ is broad, with
correlations between their values at different $k$ and $z$. We perform
a PCA to de-correlate the parameters and to extract the independent
modes that are well constrained by observations. We do not assume any
specific functional forms of $\mu(k,z)$ and $\gamma(k,z)$, and instead
pixelise them into a $m\times n$ number of pixels in the $(k,z)$ space. Both $\mu$
and $\gamma$ are linearly equally divided into $m=15$ bins in $k$
between $0 \leq k [h/{\rm Mpc}] \leq 0.3$, where the
non-linearity is mildly small. \footnote{Note that our binning in $k$ is different from the previous works of \cite{Zhao:2009fn, Hojjatietal:12}, where the $k$-binning is
  logarithmically uniform on very large scales to test whether CMB can
  constrain any $k$-modes. Since our main interest in this work is on sub-horizon
  scales, we simply set the binning linearly equal on all
  range of $k$.} For any given $k$, we have $n=15$ bins uniform in redshift in the range of $0\leq z \leq 3$, which gives us sufficient resolution to study the degeneracy between $\mu$ and $\gamma$ using a set of experiments
considered in this work. We fix $\mu=\gamma=1$ if $z>3$ simply because the experiments we shall use in the work do not have tomographic information at those redshifts, making it impossible to investigate the variation of $\mu$ or $\gamma$ at such high redshifts. Note, however, a change in the total growth from early universe to $z=3$ {\it does} have an effect to the low-$z$ growth pattern, and this was studied in details in \cite{Zhao:2009fn, Hojjatietal:12}. We do not consider this effect in this work for simplicity. Given the above mentioned pixelisation for both $\mu$ and $\gamma$,  we then study the $2m\times n$-dimensional parameter space of $\mu$ and $\gamma$.

We characterize the cosmic expansion history and initial conditions of
the universe based on the $\Lambda$CDM model with the standard set of
six cosmological parameters with Planck priors. We take Planck's
best-fit model as the fiducial model \cite{PLANCK16}: the baryon density
$\Omega_bh^2=0.022161$, the CDM density $\Omega_ch^2=0.11889$, Hubble
parameter $H_0=100h=67.77$[km/s/Mpc], the optical depth $\tau=0.0952$,
the scalar spectral index $n_s=0.9611$, and the amplitude of scalar
perturbation $\log(10^{10}A_s)=3.0973$ at $k_0=0.05[{\rm Mpc^{-1}}]$. We
assume that the universe is flat, and $\Omega_\Lambda =
1-(\Omega_c+\Omega_b)$. The dark energy equation-of-state parameter
$w$ is fixed to be $-1$.  We also consider the galaxy linear bias
$b_i$ in each tomographic and spectroscopic redshift bin (the binning is described in section~\ref{sec:3}) as free parameters with the fiducial values
set as $b(z)=\sqrt{1+z}$,  as used in
  \cite{Euclid}. The total number of the parameters that we
consider is $2 m n+6+N_b$ where $N_b$ denotes the number of the galaxy
biases of the galaxy number count and the galaxy power spectra.

In order to know the expected constraint on each parameter and the degeneracy among different parameters from future surveys, we
calculate the covariance matrix given by
\begin{align}
	C_{ij}\equiv \langle(p_i-\bar{p}_i)(p_j-\bar{p}_j)\rangle,\label{eq:11}
\end{align}
where $\bar{p}_i$ are the fiducial value of $i$-th parameter.  Because the MG parameters are correlated with each other, it is difficult to
constrain all of the parameters individually. Thus, we perform the PCA to obtain the uncorrelated parameter combinations (i.e. the eigenmodes) of $\mu(k,z)$ and $\gamma(k,z)$. Bounds on these eigenmodes can be used to forecast constraints on all kinds of MG model. The eigenmodes of $\mu(k,z)$ and $\gamma(k,z)$ are obtained by diagonalizing the covariance matrix associated with $\mu$ and $\gamma$:
\begin{align}
	{\bf C}&={\bf W}^T{\bf \Lambda}{\bf W}, \label{eq:12}\\
	\Lambda_{ij}&=\lambda_i\delta_{ij}, \label{eq:13}\\
	{\bf W}&=({\bf w}_1,{\bf w}_2,\cdots,{\bf w}_n). \label{eq:14}
\end{align}
Each eigenmode ${\bf w}_i$ represents the independent modes in $(k,z)$ space and the expected error $\sigma_i$ of each eigenmode is given by the square root of the eigenvalue $\lambda_i$. The smaller the eigenvalue means that the corresponding eigenmode is better constrained from the assumed survey.

We can also obtain the covariance matrix of $\Sigma$ from the covariance matrix of $\mu$ and $\gamma$ \cite{Hojjatietal:12}
\begin{align}
		C_{\Sigma_i\Sigma_j}&=\frac{1}{4}[\mu_i\mu_j C_{\gamma_i\gamma_j}+(1+\gamma_i)(1+\gamma_j) C_{\mu_i\mu_j}+(1+\gamma_i)\mu_jC_{\mu_i\gamma_j}+\mu_i(1+\gamma_j)C_{\gamma_i\mu_j}]. \label{eq:15}
\end{align}
We also perform a PCA of $\Sigma$ in this work.

\section{Observables}\label{sec:3}
\subsection{Measurements}
We forecast the constraints on MG parameters from lensing tomography and galaxy redshift surveys combined with CMB temperature and polarization maps.

Lensing tomographic surveys provide measurements of the weak lensing shear (WL), the angular galaxy-galaxy auto-correlation or galaxy number counts (GC), and their cross-correlation known as galaxy-galaxy lensing. The auto- and cross-correlation functions in the angular space can be written as $C^{XY}(\theta)\equiv \langle X(\hat{\bf{n}}_1)Y(\hat{\bf{n}}_2)\rangle$, where $X$ and $Y$ denote the lensing, density fluctuation and CMB temperature and polarization fields. The correlation functions can be further be expanded into the Legendre series:
\begin{align}
	C^{XY}(\theta)=\sum^\infty_{\ell=0}\frac{2\ell+1}{4\pi}C^{XY}_\ell P_\ell(\cos{\theta}), \label{eq:2-1}
\end{align}
where $C^{XY}_\ell$ is the angular power spectrum, and can be rewritten in the flat universe as
\begin{align}
	C^{XY}_\ell=4\pi\int \frac{dk}{k}\Delta^2_{\mathcal{R}}I_\ell^X(k)I_\ell^Y(k), \label{eq:2-2}
\end{align}
where $\Delta^2_{\mathcal{R}}$ is the primordial curvature power spectrum, and $I_\ell (k)$ are the transfer functions defined as
\begin{align}
	I^X_\ell(k)=\int^{z_\ast}_0dz W_X(z)j_\ell [kr(z)]\tilde{\mathcal{X}}(k,z), \label{eq:2-3}
\end{align}
where the redshift $z_\ast$ is high enough so that the standard initial conditions can be applied, $W_X(z)$ is the window function related to the redshift distribution of observables, $j_\ell$ are the Bessel function, $r(z)$ is the comoving distance, and $\tilde{\mathcal{X}}(k,z)$ is the Fourier transform of the three-dimensional field $\mathcal{X}(\hat{n}r(z),z)$.  We bin the galaxies in several photometric redshifts and write the transfer
function of GC and WL respectively as,
\begin{align}
	I^{G_i}_\ell (k)&=b_i^{\rm 2D}\int ^{z_\ast}_0 dz W_{G_i}(z)j_\ell[kr(z)]\delta(k,z),\label{eq:2-4}\\
	I^{\kappa_i}_\ell (k)&=\int^{z_\ast}_0 dz W_{\kappa_i}(z)j_\ell[kr(z)](\Psi+\Phi), \label{eq:2-5}
\end{align}
where $b_i^{\rm 2D}$ is the galaxy linear bias in the $i$-th tomographic redshift bin, $\delta(k,z)$ is the density contrast transfer function, and $W_{G_i}(z)$ is the normalized selection function for the $i$-th tomographic redshift bin given by
\begin{align}
	W_{G_i}(z)&=\frac{N_{G_i}(z)}{N^i},\label{eq:2-6}\\
	N_{G_i}(z)&=\frac{1}{2}N_G(z)\left[{\rm erfc}{\left(\frac{z_{i-1}-z}{\sqrt{2}\sigma(z)}\right)}-{\rm erfc}{\left(\frac{z_{i}-z}{\sqrt{2}\sigma(z)}\right)}\right],\label{eq:2-7}
\end{align}
where erfc is the complementary error function, $N^i$ is the total number of galaxies in the $i$-th tomographic redshift bin, and $N_G(z)$ is the angular number density of galaxies per redshift. We assume a Gaussian distribution of source galaxies around the mean redshift $z_i$ with the photometric redshift scatter of $\sigma(z)=\Delta z(1+z)$. In eq.~(\ref{eq:2-5}), $W_{\kappa_i}(z)$ denotes the window function for the $i$-th tomographic redshift bin of sheared galaxies given by
\begin{align}
	W_{\kappa_i(z)}=\int^\infty_z dz'\frac{r(z')-r(z)}{r(z)}W_{S_i}(z'),\label{eq:2-8}
\end{align}
where $W_{S_i}(z)$ is the normalized redshift distribution.  WL depends on both $\mu$ and $\gamma$, since both of them affect the lensing potential $\Phi+\Psi$.  GC probes the growth of structure and thus depends primarily on $\mu$. GC also depends on $\gamma$ via the magnification bias \cite{Turneretal:84,Villumsen:95}.

Cosmic microwave background (CMB) depends on $\Sigma$ through the Integrated Sachs-Wolfe (ISW) effect
\begin{align}
	I^{\rm ISW}_\ell(k)=\int^{z_\ast}_0dz e^{-\tau(z)}j_\ell[kr(z)]\frac{\partial}{\partial z}[\Psi+\Phi], \label{eq:2-9}
\end{align}
where $\tau(z)$ is the opaqueness function. Here we do not take into account the CMB lensing effect.

We consider the cross-power spectra among WL, GC and CMB. The cross correlation of WL with GC, i.e., galaxy-galaxy lensing not only
increases the statistical accuracy but is also important to eliminate the systematic uncertainty due to the galaxy bias. We also include the angular cross-power spectra of CMB with WL and GC generated via the ISW effect. If we divide the photometric galaxies into $M$ bins for GC and $N$ bins for WL, the total number of the angular power spectra obtained from CMB, WL, GC and their cross correlations becomes $3+M(M+1)/2+N(N+1)/2+M+N+MN$.  We assume that CMB polarization is not correlated with WL and GC measurements.

Galaxy redshift surveys provide information about the 3-dimensional distribution of galaxies and the peculiar velocities through RSD. We use the information from galaxy power spectra in the redshift-space (3D) observed from the spectroscopic surveys
\begin{align}
	P^{\rm obs}_g(k,c,z)=P_{gg}(k,z)+2c^2P_{g\Theta}(k,z)+c^4P_{\Theta\Theta}(k,z),\label{eq:2-10}
\end{align}
where $c$ is the cosine of the angle between ${\bf k}$ and the line of sight, $P_{gg}$ is the true galaxy power spectrum, and $P_{\Theta\Theta}$ is the power spectrum of the normalized peculiar velocity $\Theta\equiv \theta/aH$, and $P_{g\Theta}$ is the galaxy-velocity cross spectrum.  We forecast the constraints on MG by calculating the angular power spectra and the matter power spectra using MGCAMB \cite{MGCAMB,G.B.Zhaoetal:09,Hojjati:2011ix}.

\subsection{Fisher analysis}
In order to estimate the uncertainty in each parameter, we perform
the Fisher analysis (see \cite{G.B.Zhaoetal:09,Hojjatietal:12} for details).
According to the Cram\'er-Rao inequality, the inverse of the Fisher matrix
gives the lower bound on the variance in a given parameter $p_i$ as $1/F_{ii}$ (other
  parameters fixed) or $F^{-1}_{ii}$ (other parameters marginalised
  over),
\begin{equation}
F_{ab}^{\rm 2D}=f_{\rm sky}\sum^{\ell_{\rm max}}_{\ell=\ell_{\rm min}}
\frac{2\ell+1}{2}{\rm Tr}\left(\frac{\partial {\bf C}_\ell}{\partial p_a}
\tilde{\bf C}_\ell^{-1}\frac{\partial {\bf C}_\ell}{\partial p_b}\
\tilde{\bf C}_\ell^{-1}\right),
\label{eq:2-11}
\end{equation}
where $p_a$ is the $a$-th parameter and $\tilde{\bf C}_\ell$ is the
covariance matrix of the angular power spectra containing the noise
\cite{Tegmarketal:97}. The range of $\ell$ is determined as $\ell_{\rm
  min}\simeq \pi/(2f_{\rm sky})$, assuming the observed sky is
contiguous, and $\ell_{\rm max}=2000$ to exclude the non-linear regime
\cite{Kitching:2010ab}.
Varying $\mu$ or $\gamma$ at a certain scale $k$ mainly
  affects $C_\ell$ with the corresponding angular scale $\ell=k r(z)$,
  where $r(z)$ is the comoving distance.
The elements of $\tilde{\bf C}_\ell$ are given by
\begin{align}
	\tilde{C}^{XY}_\ell=C^{XY}_\ell+N^{XY}_\ell, \label{eq:2-12}
\end{align}
where $X$ and $Y$ denote GC or WL at some tomographic redshift bin. Eq.~(\ref{eq:2-11}) can be rewritten as
\begin{equation}
F_{ab}^{2D}=f_{\rm sky}\sum^{\ell_{\rm max}}_{\ell=\ell_{\rm min}}(2\ell+1)
\sum_{ij}\sum_{mn}\frac{\partial C_\ell^{X_iX_j}}{\partial p_a}
{\bf {\cal C}}_\ell^{-1}\frac{\partial C_\ell^{X_mX_n}}{\partial p_b},
\end{equation}
and the elements of the covariance matrix are
\begin{equation}
{\cal C}_\ell^{ij,mn}=\tilde{C}_\ell^{X_iX_m}\tilde{C}_\ell^{X_jX_n}+
\tilde{C}_\ell^{X_iX_n}\tilde{C}_\ell^{X_jX_m}.
\end{equation}
We only consider the statistical errors of the GC and WL auto correlations at same tomographic redshift bins
\begin{align}
	N_\ell^{\kappa_i\kappa_j}&=\delta_{ij}\frac{\gamma_{\rm int}^2}{n^i_{\rm gal}},\\
	N_\ell^{G_iG_j}&=\delta_{ij}\frac{1}{n^i_{\rm gal}},\\
	N_\ell^{\kappa_iG_j}&=0,
\end{align}
where $n_{\rm gal}^{i}$ is the angular number density per steradian in the $i$-th tomographic redshift bin and $\gamma_{\rm int}$ is the intrinsic ellipticity of galaxies.
We numerically compute the derivatives $\partial C_\ell^{XY}/\partial p_a$ using MGCAMB by slightly shifting each parameter from its fiducial value. For simplicity, we neglect various observational systematics such as photometric redshift errors and shape measurement error. See ref.~\cite{Hojjatietal:12} for a study of the influence of these systematics.
The fraction of the contiguous sky area $f_{\rm sky}$ depends on the measurement $X(\hat{\bf n})$ and also on the assumed surveys.

The Fisher matrix for the galaxy power spectra in the redshift-space is given by \cite{Tegmark:97},
\begin{align}
	F_{ab}^{\rm 3D}=\sum_i\int_0^{k_{\rm max}} \frac{k^2dk}{2(2\pi)^2}\int ^{1}_{-1}dc V_{\rm eff}(k,c,z_i)\frac{\partial \ln{P^{\rm obs}_g(k,c,z_i)}}{\partial p_a}\frac{\partial \ln{P^{\rm obs}_g(k,c,z_i)}}{\partial p_b}, \label{eq:2-13}
\end{align}
where $k_{\rm max}$ is set to be $0.35[h/{\rm Mpc}]$ and $V_{\rm eff}(k,c,z_i)$ is the effective volume in each spectroscopic redshift bin given by
\begin{align}
	V_{\rm eff}(k,c,z_i)&=\left[\frac{n^i_{\rm 3D}P^{\rm obs}_g(k,c,z_i)}{n^i_{\rm 3D}P^{\rm obs}_g(k,c,z_i)+1}\right]^2V_{\rm survey}(z_i), \label{eq:2-14}\\
	V_{\rm survey}(z_i)&=\frac{4\pi}{3}f_{\rm sky}[r(z_i+\Delta z/2)^3-r(z_i-\Delta z/2)^3],\label{eq:2-15}
\end{align}
and $n^i_{\rm 3D}$ is the number density of the galaxies in each spectroscopic redshift bin. The numbers densities of the galaxies in the spectroscopic redshift bins are listed in table~\ref{ta:1}. We do not take into account the covariance between different spectroscopic redshift bins.  We use a simple Kaiser formula to describe the galaxy power spectra $P_{\rm gg}$, the peculiar velocity power spectra $P_{g\Theta}$, and the galaxy-velocity cross spectra in the redshift-space $P_{\Theta\Theta}$:
\begin{align}
	P_{gg}(k,z)&=(b_i^{\rm 3D})^2P_{\delta\delta}(k,z),\label{eq:2-16}\\
	P_{\Theta\Theta}(k,z)&=f^2P_{\delta\delta}(k,z),\label{eq:2-17}\\
	P_{g\Theta}(k,z)&=\sqrt{P_{gg}P_{\Theta\Theta}},\label{eq:2-18}
\end{align}
where $b_i^{\rm 3D}$ is the galaxy linear bias in the $i$-th spectroscopic redshift bin, independent of $b_i^{\rm 2D}$ in eq.~(\ref{eq:2-4}), $f\equiv d\ln{D}/d\ln{a}$ is the logarithmic derivative of the growth factor, and $P_{\delta\delta}$ is the matter power spectrum.  We assume that the effect of $\gamma$ on the growth of matter is negligibly small, $\partial P^{\rm obs}_g/\partial\gamma=0$. Here we treat 2D and 3D measurements independently and simply sum up their Fisher values. Actually, there is correlation between the 2D and 3D measurements when they cover the same area.
As discussed in Subsection 2.5 in \cite{Gaztanagaetal:12}, such correlation depends on k and decreases as the number of independent modes in the radial direction
increase. Therefore, rather than using only the radial mode information from the 3D spectra, we are using all of the information including the transverse modes,  as it is negligible compared to the 2D. The correlation can be quantitatively estimated using numerical simulations, however, the detailed analysis is left for
future works.

\subsection{Experiments}
As in \cite{G.B.Zhaoetal:09}, we assumed CMB data from the three lowest frequency HFI channels of Planck with $f_{\rm sky}=0.8$. We also assume tomographic galaxy catalogues and WL data from a Euclid-like \cite{Euclid} survey, as well as spectroscopic galaxy catalogues from a Euclid-like ($0.65\leq z \leq 2.05$) and from BOSS-like ($0.35\leq z \leq 0.65$) \cite{BOSS} surveys. The redshift distribution of photometric galaxies in the tomography survey is given by
\begin{align}
	N_G(z)\propto z^2\exp{(-(z/z_0)^{3/2})},\label{eq:2-19}
\end{align}
where $z_0=z_{\rm mean}/1.412$ is the peak of $N_G(z)$ and we assume
that the median redshift is $z_{\rm mean}=0.9$, the surface galaxy
number density is $n_{\rm gal}=30\ {\rm per\ arcmin^2}$, and the
covered region of the sky is 15,000 square degrees for both
BOSS-like \footnote{Note that the parameters we use for
    BOSS-like survey are different from those used in the actual BOSS survey, where
    the planed sky coverage is $\sim 10,000{\rm deg}^2$ and the
    redshift distribution of CMASS sample is given in figure~4 in
    \cite{Andersonetal:12}. Our choice of the number density is based
    on the BOSS project paper \cite{BOSS_project}.}
and the Euclid-like survey \cite{Euclid}.  We use the photometric
redshift error $\Delta z=0.05$, and the intrinsic galaxy shear
$\gamma_{\rm int}=0.22$. We choose to have $8$ tomographic redshift
bins for WL and GC, with the widths of the bins being wider at higher $z$ as the
photometric redshift error increases,
i.e. $z_i=(0.2,0.45,0.75,1.1,1.5,1.95,2.45,3.1)$. Table~\ref{ta:1}
lists the galaxy number densities $n^i_{\rm 3D}$ for the
BOSS-like and Euclid-like galaxy redshift surveys used in our
forecast. The number of parameters for bias $N_b$ is 8(2D) + 17(3D).
\begin{table}[tbp]
\centering
\begin{tabular}{|c|ccccccccc|}\hline
 Survey &  \multicolumn{3}{c|}{BOSS-like} &  \multicolumn{6}{c|}{Euclid-like} \\
\hline
$z$ &  0.4 & 0.5 & \multicolumn{1}{c|}{0.6} & 0.7 & 0.8 & 0.9 & 1.0 & 1.1 & 1.2  \\ \hline
$n(z) [10^{-3}(h/{\rm Mpc})^3]$ & 0.3 & 0.3 & \multicolumn{1}{c|}{0.3} & 1.25 & 1.92 & 1.83 & 1.68 & 1.51 & 1.35  \\ \hline\hline
Survey &  \multicolumn{8}{c}{Euclid-like} &\\ \hline
$z$ & 1.3 & 1.4 & 1.5 & 1.6 & 1.7 & 1.8 & 1.9 & 2.0 & \\ \hline
$n(z) [10^{-3}(h/{\rm Mpc})^3]$ & 1.20 & 1.00 & 0.80 & 0.58 & 0.38 & 0.35 & 0.21 & 0.11 & \\
\hline
\end{tabular}
\caption{Redshift survey parameters considered in this paper. Mean
  redshift of and the average galaxy number densities in each spectroscopic bins
  assumed for BOSS-like$^{2}$ and Euclid-like surveys.}
\label{ta:1}
\end{table}

\section{Results}\label{sec:4}
\begin{table}[tpb]
  \centering
  \begin{tabular} {|c|p{4.2em}p{4.2em}p{4.2em}p{4.2em}p{4.2em}p{4.2em}|} \hline
parameter	&	\multicolumn{1}{c}{$\Omega_bh^2$}	&	\multicolumn{1}{c}{$\Omega_ch^2$}	&	\multicolumn{1}{c}{$h$}	 &	\multicolumn{1}{c}{$\tau$}	&	\multicolumn{1}{c}{$n_s$}	&	\multicolumn{1}{c|}{$10^9A_s$} \\ \hline
CMB	&	$1.4\times10^{-4}$	&	$1.4\times10^{-3}$	&	$6.4\times10^{-1}$	&	$4.6\times10^{-3}$	&	 $3.7\times10^{-3}$	&	$2.0\times10^{-2}$ \\
CMB+WL	&	$1.0\times10^{-4}$	&	$3.0\times10^{-4}$	&	$1.4\times10^{-1}$	&	$4.2\times10^{-3}$	&	 $2.3\times10^{-3}$	&	$1.7\times10^{-2}$ \\
CMB+GC	&	$9.8\times10^{-5}$	&	$4.1\times10^{-4}$	&	$1.8\times10^{-1}$	&	$4.3\times10^{-3}$	&	 $2.0\times10^{-3}$	&	$1.8\times10^{-2}$ \\
CMB+WL+GC	&	$9.8\times10^{-5}$	&	$2.9\times10^{-4}$	&	$1.3\times10^{-1}$	&	$3.9\times10^{-3}$	&	 $1.5\times10^{-3}$	&	$1.6\times10^{-2}$ \\
CMB+3D	&	$9.1\times10^{-5}$	&	$3.5\times10^{-4}$	&	$1.8\times10^{-1}$	&	$2.6\times10^{-3}$	&	 $2.3\times10^{-3}$	&	$1.2\times10^{-2}$ \\
2D	&	$9.7\times10^{-5}$	&	$2.5\times10^{-4}$	&	$1.1\times10^{-1}$	&	$3.2\times10^{-3}$	&	 $1.5\times10^{-3}$	&	$1.3\times10^{-2}$ \\
2D+3D	&	$8.3\times10^{-5}$	&	$2.0\times10^{-4}$	&	$6.2\times10^{-2}$	&	$2.0\times10^{-3}$	&	 $1.2\times10^{-3}$	&	$7.4\times10^{-3}$ \\
2D+3D (MG)	&	$9.0\times10^{-5}$	&	$2.3\times10^{-4}$	&	$7.4\times10^{-2}$	&	$2.7\times10^{-3}$	&	 $1.6\times10^{-3}$	&	$1.1\times10^{-2}$ \\ \hline
    \end{tabular}
\caption{The forecasted $1\sigma$ uncertainties in the cosmological parameters expected from CMB (Planck) only and its combined with WL, GC and 3D  information respectively. 2D means all information obtained by combining CMB, WL, and GC auto- and cross-power spectra.  MG means the results after marginalizing over the MG parameters, i.e. $\mu$ and $\gamma$. Galaxy bias parameters in GC and 3D measurements are marginalized over for all of the results.}
    \label{ta:2}
  \end{table}

\begin{figure}[tpb]
  \centering
    \includegraphics[width=.65\textwidth]{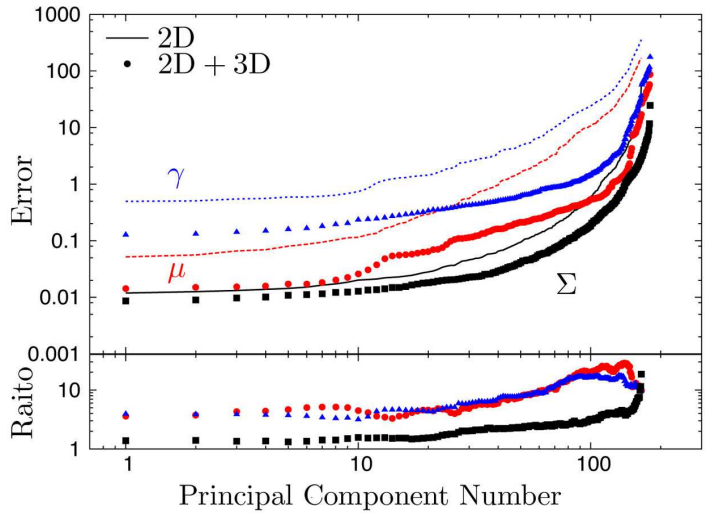}
  \caption{The upper panels show the uncertainties in the amplitudes of the principal components
    (eigenmodes) of $\mu$ (red circles), $\gamma$ (blue triangles) and
    $\Sigma$ (black squares) from 2D+3D information. The number of
    principal components is ordered from small to large $\sigma$. For
    comparison, the lines shows the corresponding errors obtained from the 2D information
    only. The lower panel shows the ratio of errors for the eigenmodes
    with the same order number, which indicates the extent to which the
    eigenmodes are better constrained by adding the 3D
    information. Cosmological parameters and the galaxy bias
    parameters are marginalized over.}
  \label{fig:2}
\end{figure}

In this section we show the results of the principal component analysis of $\mu$, $\gamma$ and $\Sigma$ from 2D measurements (CMB, WL and GC auto- and cross-spectra) combined with 3D measurements of the galaxy redshift-space power spectra.  We particularly focus on how the 3D information improves the constraints on the eigenmodes of $\mu$, $\gamma$ and $\Sigma$.

\begin{figure}[tpb]
\centering
	\includegraphics[width=.45\textwidth]{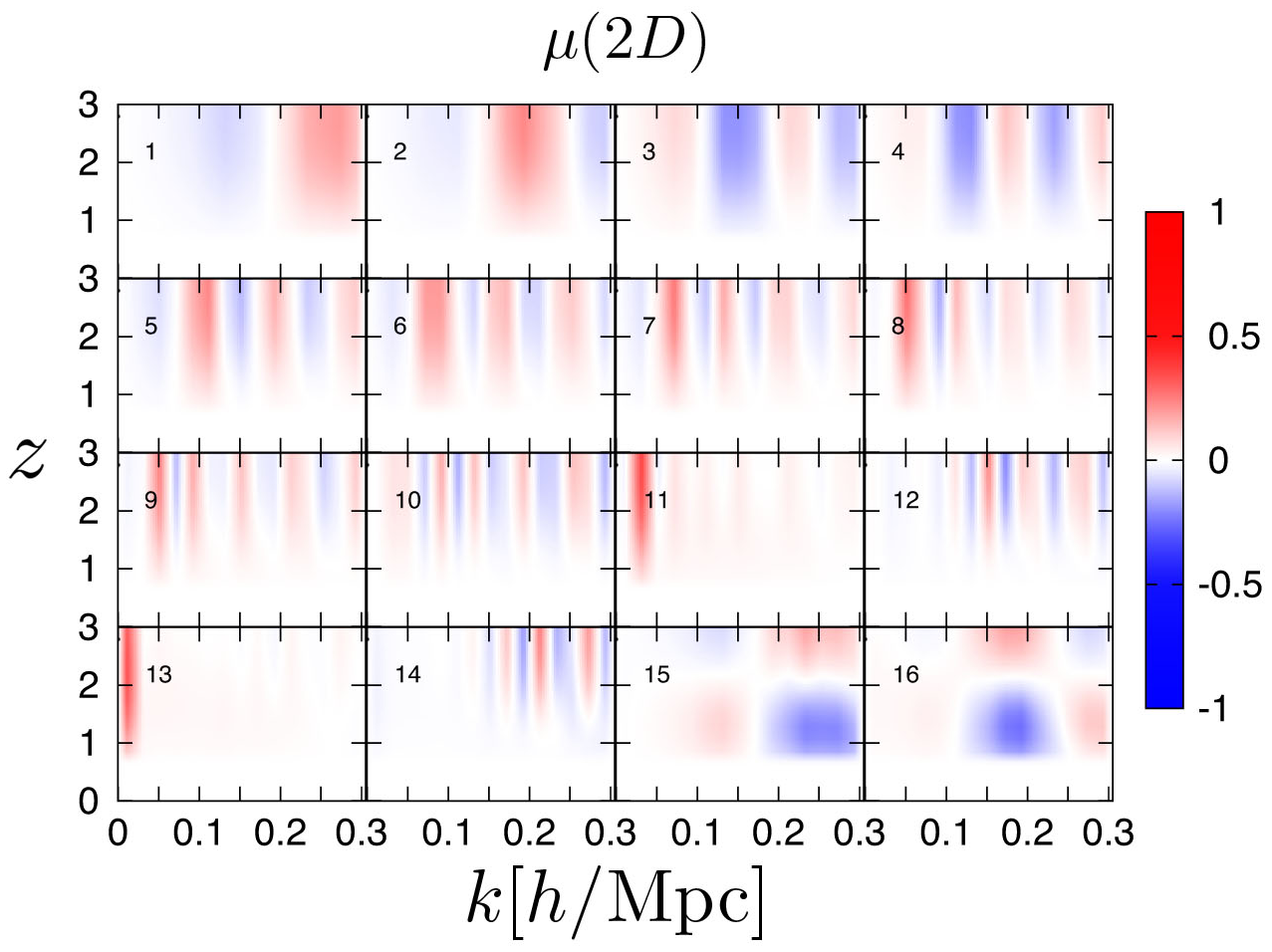}
\hfill
\centering
	\includegraphics[width=.45\textwidth]{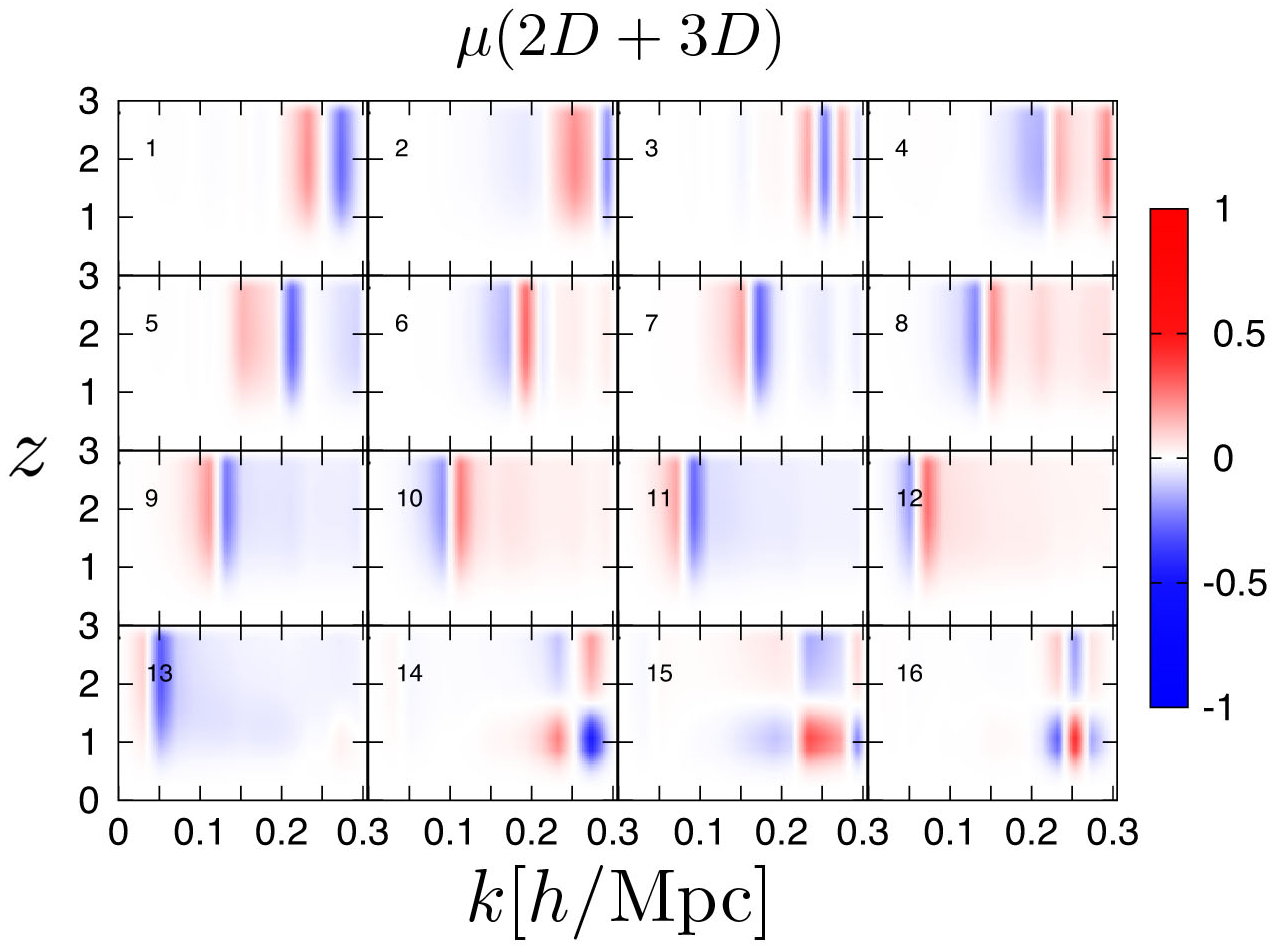}
\newline
\centering
	\includegraphics[width=.45\textwidth]{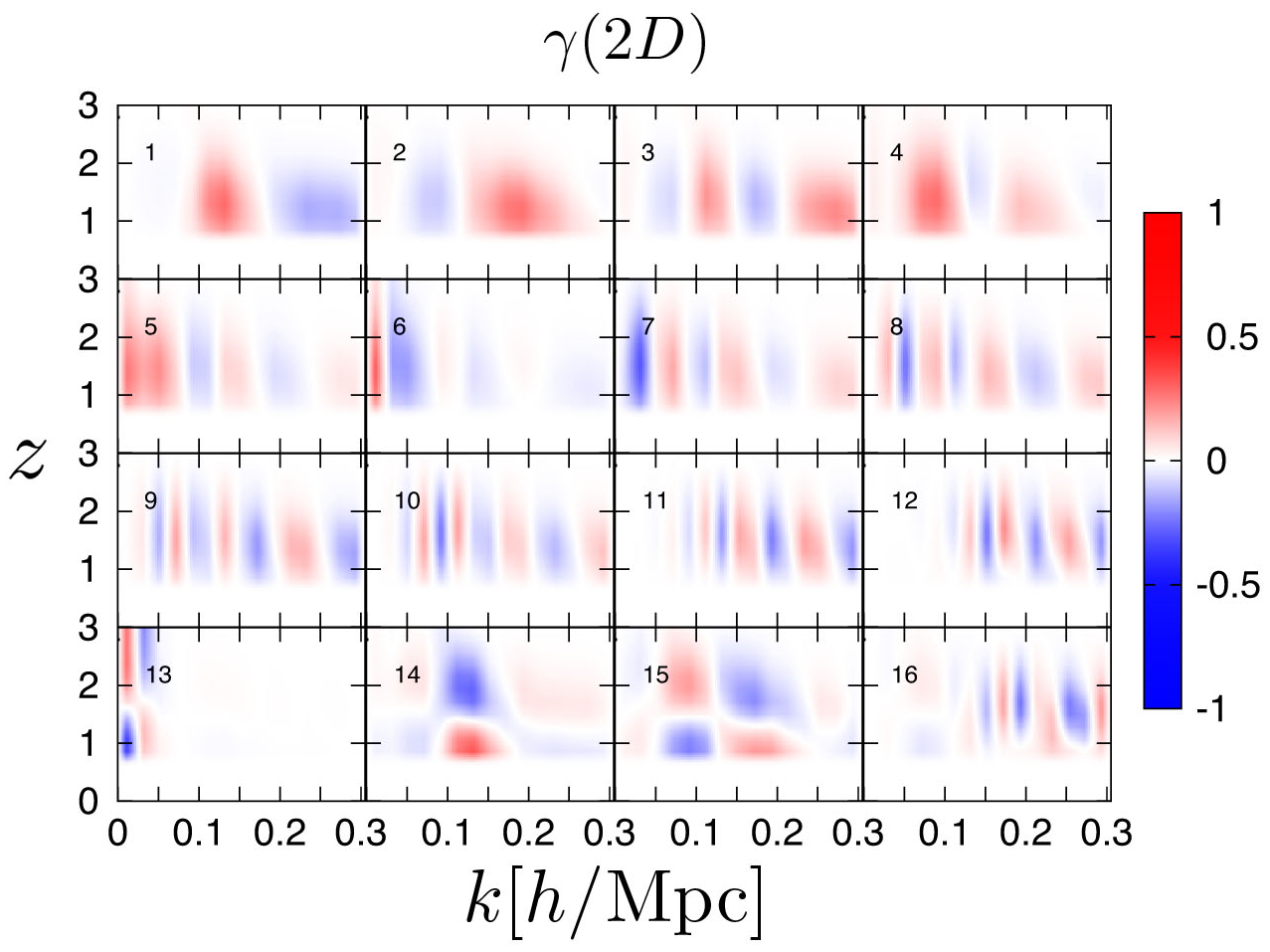}
\hfill
\centering
	\includegraphics[width=.45\textwidth]{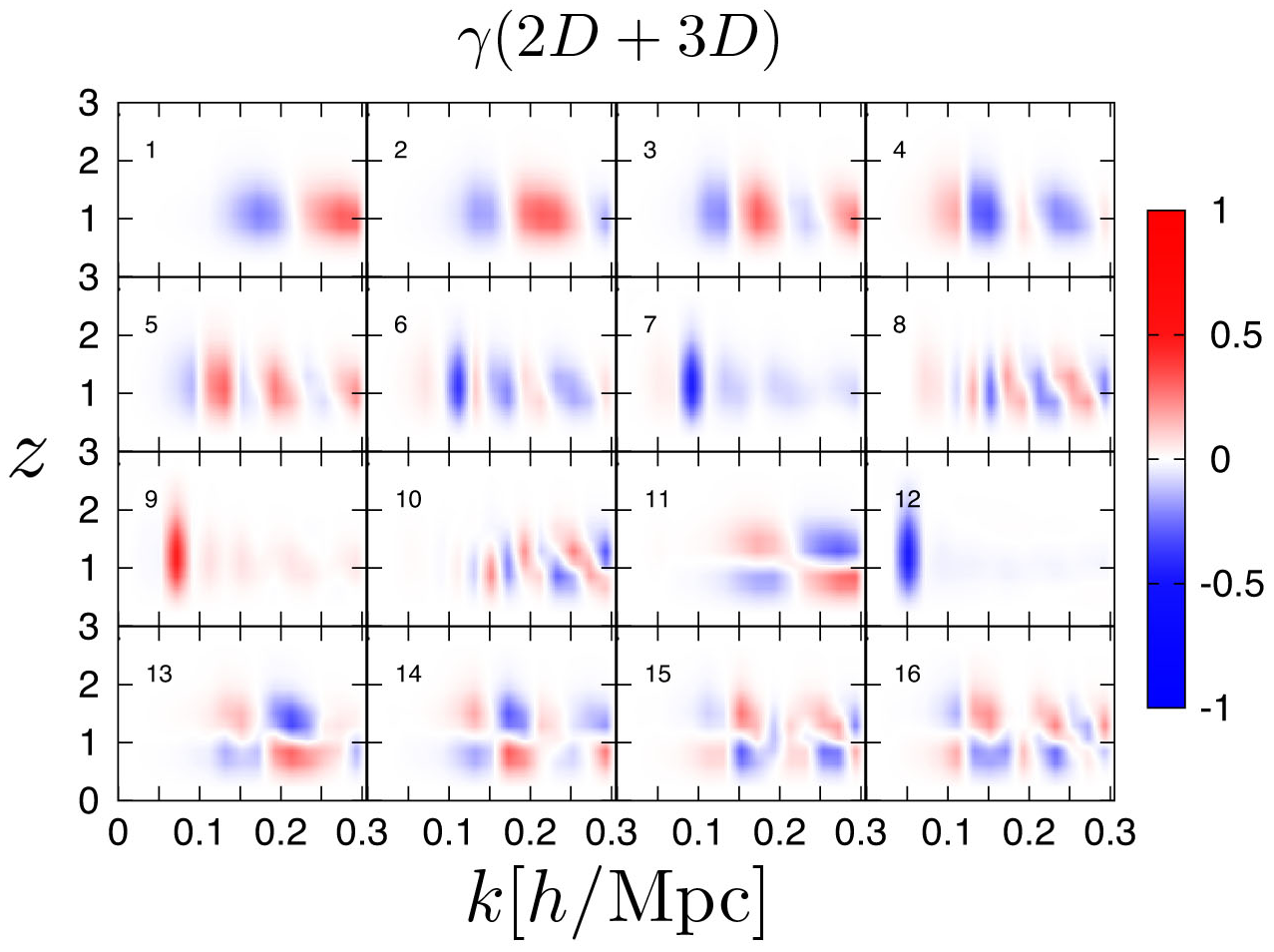}
\newline
\centering
	\includegraphics[width=.45\textwidth]{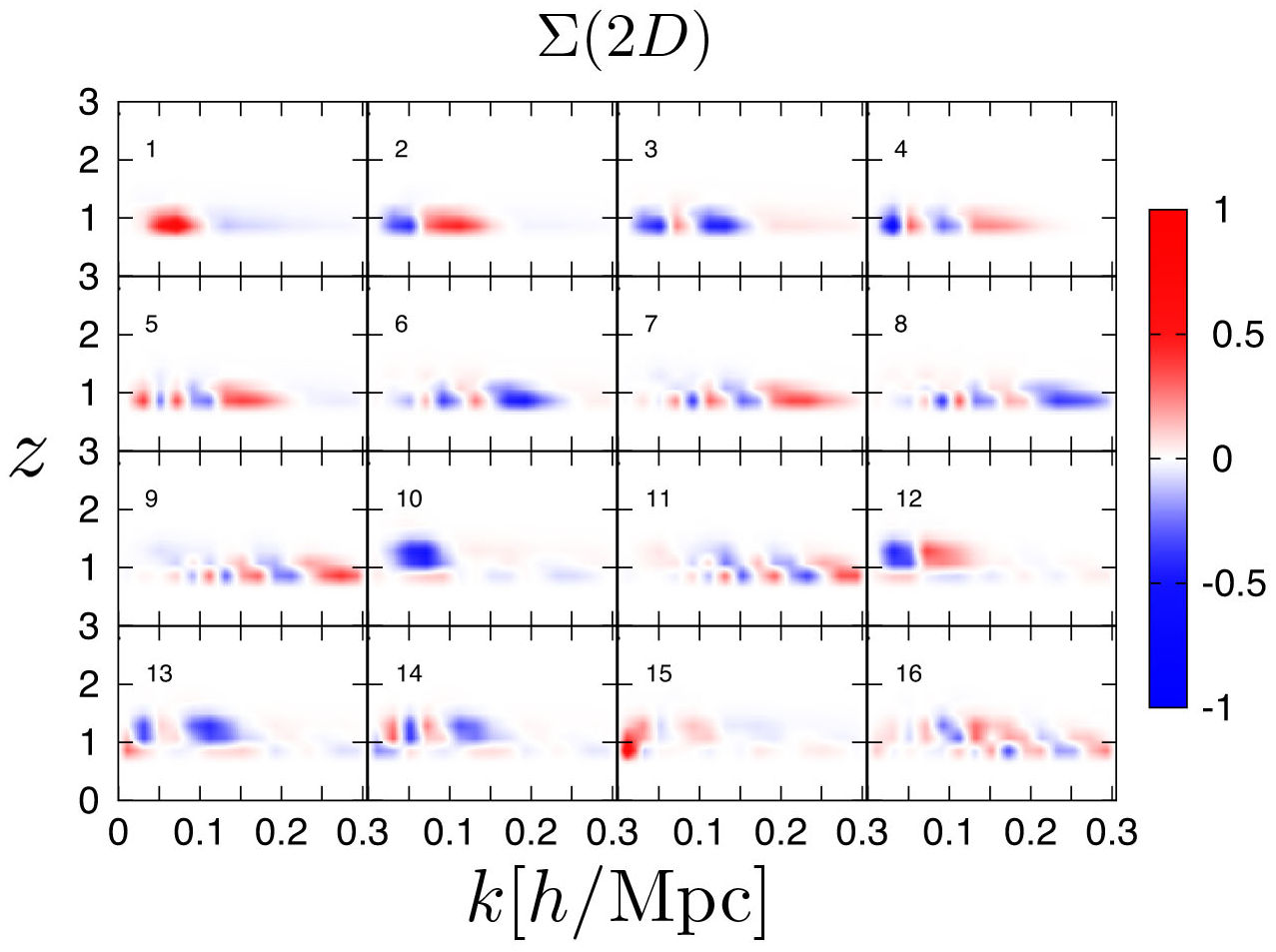}
\hfill
\centering
	\includegraphics[width=.45\textwidth]{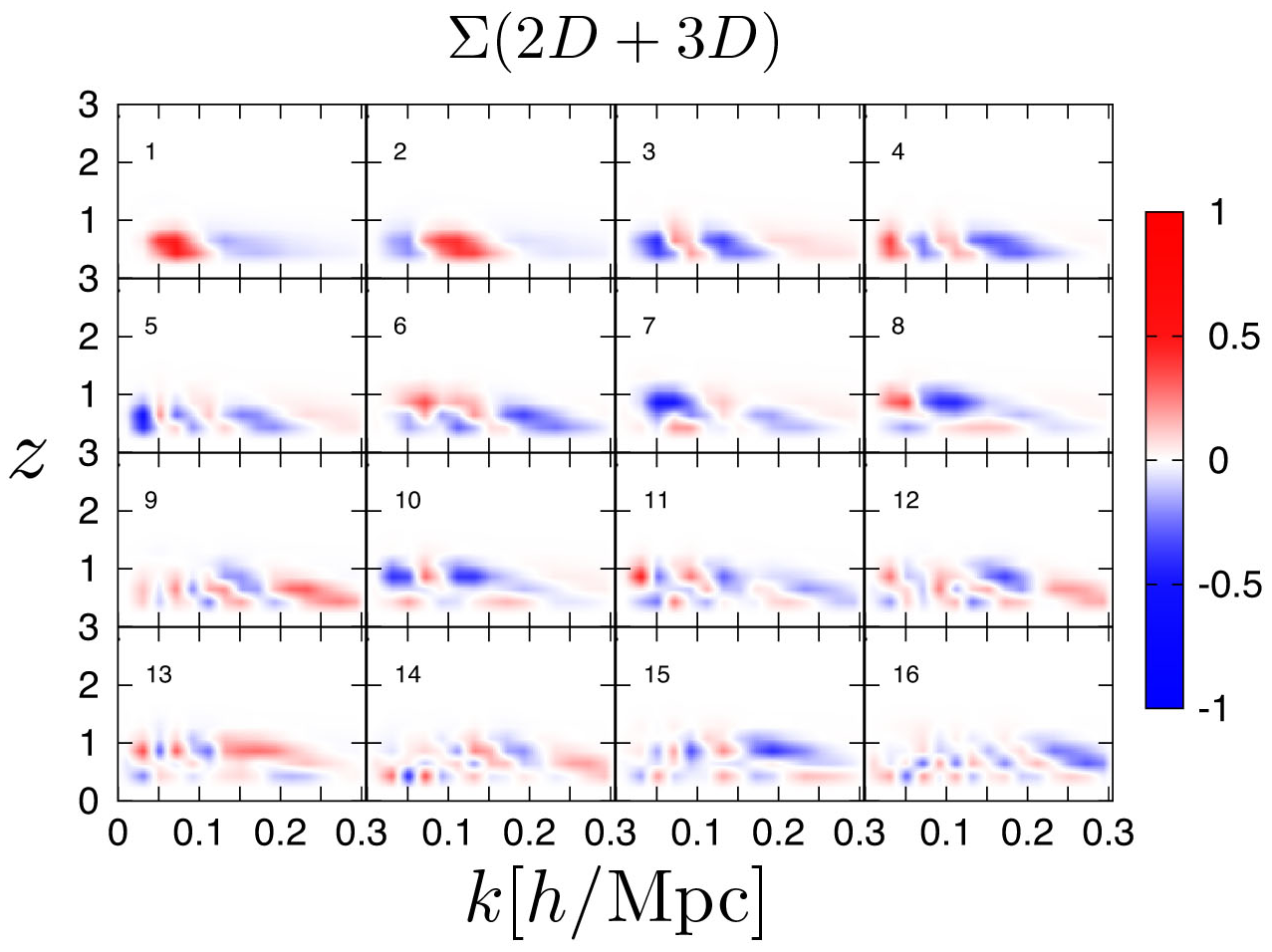}
  \caption{Eigenmodes of $\mu$, $\gamma$, and $\Sigma$ from top to bottom.
    The left panels show the eigenmodes calculated from the 2D information (WL$\times$GC$\times$CMB)
    and the right panels are those from the 2D+3D information.}
  \label{fig:3}
\end{figure}

First we investigate the degeneracy of MG parameters with $6$ cosmological parameters and the bias parameters. Table~\ref{ta:2} lists the forecasted uncertainties in the cosmological parameters from the combinations of the various observations. \footnote{We checked that our forecasts for the constraints on the cosmological parameters from CMB are consistent with the result of \cite{Colomboetal:08}. Note that the actual constraints on the cosmological parameters from Planck \cite{PLANCK16} are worse, because the polarization data has not been included in the Planck analysis yet, but it is included in our forecast.}
The cosmological parameters are constrained mainly by CMB data, while large-scale structure measurements significantly improve the accuracy of $\Omega_ch^2$ and $h$ because they are sensitive to $\Omega_c$.  When adding $\mu(k,z)$ and $\gamma(k,z)$ as free parameters, the marginalized constraints on cosmological parameters weaken by up to 50\% mainly due to the strong degeneracy between the amplitude $A_s$ and the MG parameters: the change of the initial fluctuation amplitude can be compensated by the overall change of $\mu$. We also find that the error on bias parameters in tomographic and spectroscopic bins increases only by about 10\% after adding MG parameters. The degeneracy of the galaxy bias with MG parameters is small because we use the linear bias that does not depend on scale and almost all the well-constrained eigenmodes in figure~\ref{fig:3} are oscillating along $k$-axis. Degeneracies between the cosmological parameters and different sets of MG parameters have been studied in detail in \cite{Hojjati:2012ci}.

The upper panel of figure~\ref{fig:2} shows the forecasted errors on amplitudes of principal components (eigenmodes) in the ascending order of $\sigma$, i.e. the eigenmodes are ordered from best constrained to worst. The lines represent the errors on eigenmodes of $\mu$, $\gamma$ and $\Sigma$ from 2D only, while symbols represent the corresponding errors from a combination of 2D with 3D. In all cases, the covariance matrix is estimated by taking the inverse of a $(2mn+6+N_b)\times(2mn+6+N_b)$ Fisher matrix, but the eigenmodes and eigenvalues are estimated by separately diagonalizing the $mn\times mn$ covariance submatrices associated with $\mu$ and $\gamma$. The covariance matrix of $\Sigma$ is calculated by using eq.~(\ref{eq:15}). The lower panel of figure~\ref{fig:2} represents the ratio of the errors of eigenmodes with the same principal component number between 2D and 2D+3D. We find that adding 3D reduces the errors of $\mu$ and $\gamma$ by a factor between $3$ and $10$.

Note that the plots in figure~\ref{fig:2} just compare the eigenvalues with and without 3D information.
They do not provide any information about the amplitude of the corresponding modes in modified gravity theories.
For example, for some of these theories,  departure from the fiducial value of $\Sigma$ (and hence the amplitude of the oscillating modes) is small. The Fisher matrices for parameters in specific MG models can be calculated by 
projecting errors on the parameters from principal components without regenerating the Fisher matrix 
from scratch \cite{Zhao:2009fn,Hojjatietal:12}. 
This projection was done for a one-parameter model of $\mu(k,a)$ and $\gamma(k,a)$ which gives a good
approximations for $f(R)$ theories in quasi-static limits in Ref.~\cite{Hojjatietal:12} using only 2D information. We leave detailed studies of the effects of adding 3D information on constraining parameters in specific MG models for future work.


Note that we used different redshift ranges for 2D only and 2D+3D measurements: $0.8 \leq z \leq 3.0$ for 2D only and $0.4 \leq z \leq 2.8$ for 2D+3D.
Here we consider the spectroscopic samples where galaxies populate at the redshift from 0.4 to 2 and then MG parameters are strongly degenerate outside this redshift range.
Accordingly, the number of $z$-bins changes from $n=11$ (2D) to $n=12$ (2D+3D) with the binning width fixed to be 0.2 in redshift. Even though the maximum redshift for 2D+3D is smaller than that for 2D only, the result is insensitive to the maximum redshift because the number of observed galaxies at high redshift is small. A quantitative study on the effect of varying $\mu$ and $\gamma$ at high-$z$ is left for future work.
A significant result of adding the 3D information is that the number of eigenmodes with $\sigma<1$ increases from $40$ to $123$ for $\mu$, from $11$ to $81$ for $\gamma$ and from $110$ to $140$ for $\Sigma$ after adding the 3D information. This improvement comes from the additional information in the radial modes provided by the 3D data. For $\Sigma$, eigenmodes are primarily constrained from WL observables and thus the improvement from adding the 3D information is relatively small.

Figure~\ref{fig:3} shows the first 16 principal component modes (eigenmodes) with small uncertainties for $\mu$, $\gamma$ and $\Sigma$, from top to bottom, in 2D (left) and 2D+3D (right). Each mini-panel describes ${\bf w}_i$ (eq.~(\ref{eq:14})) in the $(k,z)$ space (colours represent the amplitude).  As shown in \cite{Hojjatietal:12}, if only the 2D measurements are used, the best constrained eigenmodes show oscillations in $k$, while new $z$-oscillation modes appear after each dozen or so $k$-oscillation modes. This indicates that the scale dependence of $\mu$ and $\gamma$ is much
better constrained than the $z$-dependence. This is because the departure of $\mu$ from unity at a certain $z_i$ affects the clustering at all lower $z$ and thus the degeneracy along $z$-axis becomes strong. Also, the WL kernel for a given angular moment $\ell$ receives contributions from $k$ and $z$ over a relatively wide range, and directly probes $\Sigma$ rather than $\mu$ and $\gamma$ individually. This results in an additional loss of sensitivity to the dependence of $\mu$ on $z$.
On the other hand, when the 3D measurements are added, each eigenmode becomes sensitive to $\mu$ at a certain $k$ because the departure of $\mu$ from unity at a certain scale $k$ affects the 3D galaxy power spectrum only at the corresponding scale in the linear approximation. This is the reason why the eigenmodes of $\mu$ after adding the 3D measurements have peaks around a certain $k$. Some of the eigenmodes of $\gamma$ (e.g. the $7$-th, $9$-th and $12$-th) are also sensitive to specific scales, but the other eigenmodes still show oscillations in $k$ because $\gamma$ is constrained by both the angular power spectra and the galaxy power spectra. On the other hand, $\Sigma$ is mainly determined by the 2D measurements thus their eigenmodes do not change much even if we add the 3D information.

\section{Degeneracies between $\mu$ and $\gamma$ }\label{sec:5}

In order to understand the cause of the degeneracy between $\mu$ and $\gamma$, and how the 3D information resolves it, we consider the simplest case where the MG parameters consist of just 2 parameters $\mu$ and $\gamma$ at some specific $k$ and $z$ and neglect the correlation with other $k$ and $z$. We also ignore the degeneracy with other parameters, such as  $\Omega_M$ and the bias, here for simplicity.

\begin{figure}[tbp]
\centering
	\includegraphics[width=.45\textwidth]{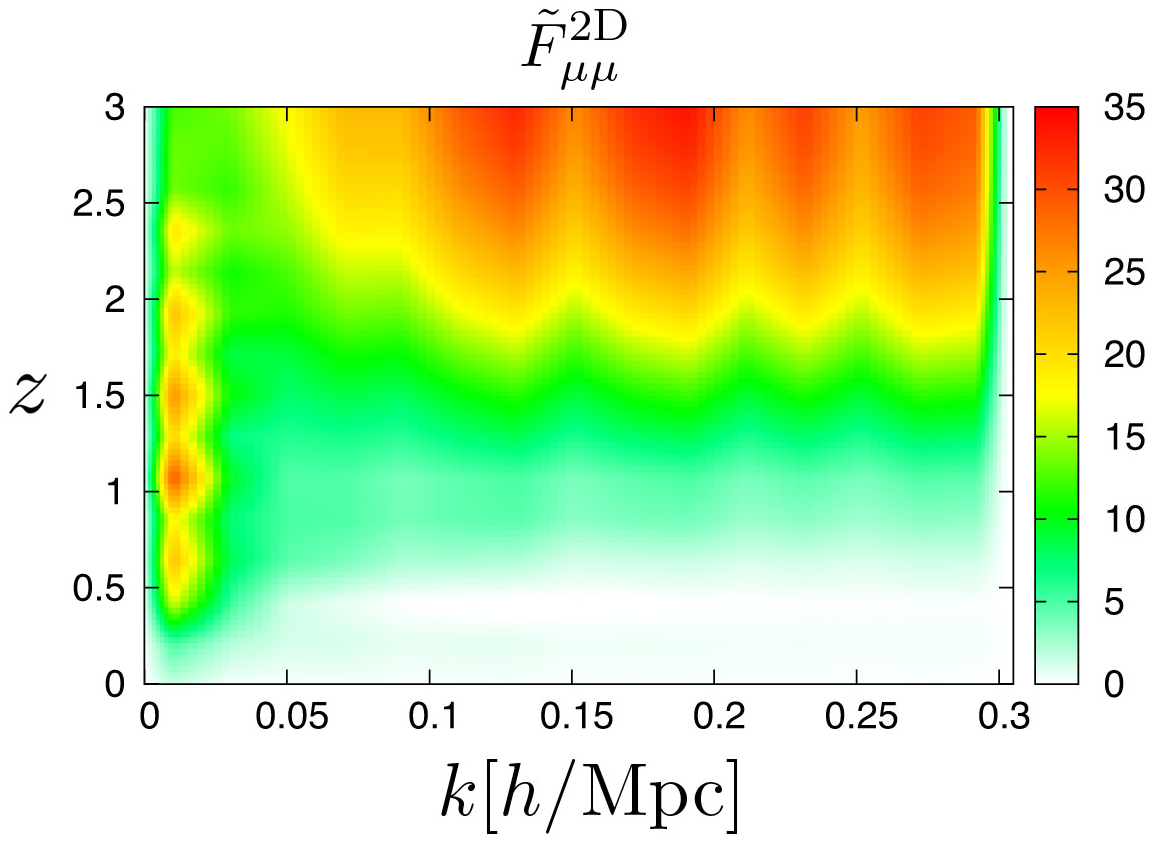}
\hfill
\centering
	\includegraphics[width=.45\textwidth]{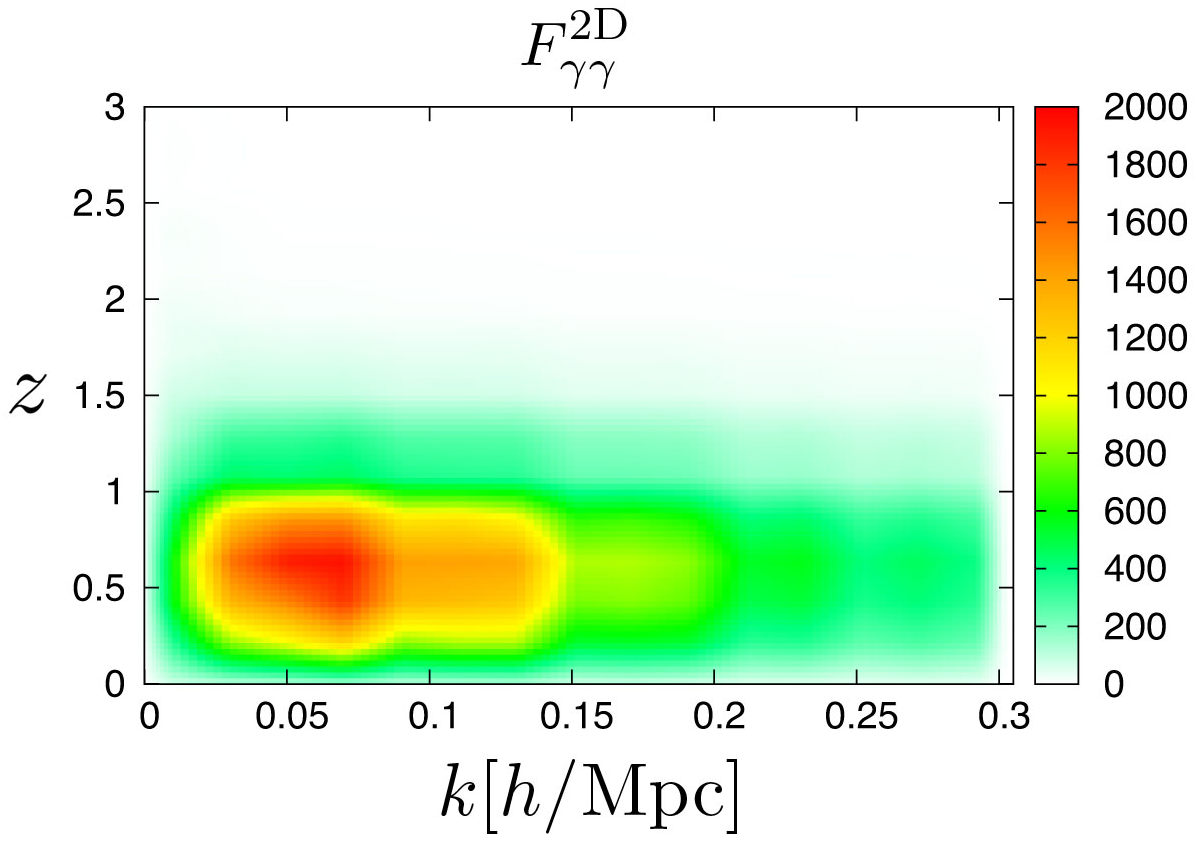}
  \caption{Left and right panels show the values of $\tilde{F}_{\mu\mu}$ and\
    $F_{\gamma\gamma}$ representing the inverse of the variance of $\mu$
    and $\Sigma$ (see eqs.~(\ref{eq:5-5b}) and (\ref{eq:5-6})),
    respectively, at different $k$ and $z$.}
  \label{fig:5-1}
\end{figure}

First we consider the 2D measurements only. From the relation between
the lensing potential and the matter perturbation, eq.~(\ref{eq:9}),
\begin{align}
	C_\ell^{\rm WL}\propto [\mu(1+\gamma)/2]^2,
\end{align}
and the derivatives of the angular power spectra with respect to $\mu$ and $\gamma$ are given by
\begin{align}
	\frac{\partial C^{\rm WL}_\ell}{\partial \mu}&\propto \mu(1+\gamma)^2/2,\\
	\frac{\partial C^{\rm WL}_\ell}{\partial \gamma}&\propto \mu^2(1+\gamma)/2.
\end{align}
Therefore, the response of $\mu$ and $\gamma$ to the angular power spectra of WL
has the following relation:
\begin{align}
	\frac{\partial C^{\rm WL}_\ell}{\partial \mu}=\frac{(1+\gamma)}{\mu}\frac{\partial C^{\rm WL}_\ell}{\partial \gamma}=2\frac{\partial C^{\rm WL}_\ell}{\partial \gamma}, \label{eq:5-2}
\end{align}
where we used the fiducial values of $\mu$ and $\gamma$, $\mu=\gamma=1$.
Thus the Fisher information given by WL has the relation of $F^{\rm
  WL}_{\mu\mu}=4F^{\rm WL}_{\gamma\gamma}$, and, thereby, the
parameters of $\mu$ and $\gamma$ are completely degenerate with each
other. Note that the lensing potential has additional $\mu$-dependence through the growth of the density perturbation $\delta$ but this additional $\mu$-dependence is weak and also this does not change the fact that $\mu$ and $\gamma$ are completely degenerate in WL measurements. We found that the dominant contribution of the 2D measurements comes from WL auto-correlations and cross-correlations between WL and GC through magnification bias effects and this is the origin of the degeneracy between $\mu$ and $\gamma$ when only the 2D measurements are used.

\begin{figure}[tbp]
	\centering
	\includegraphics[width=.45\textwidth]{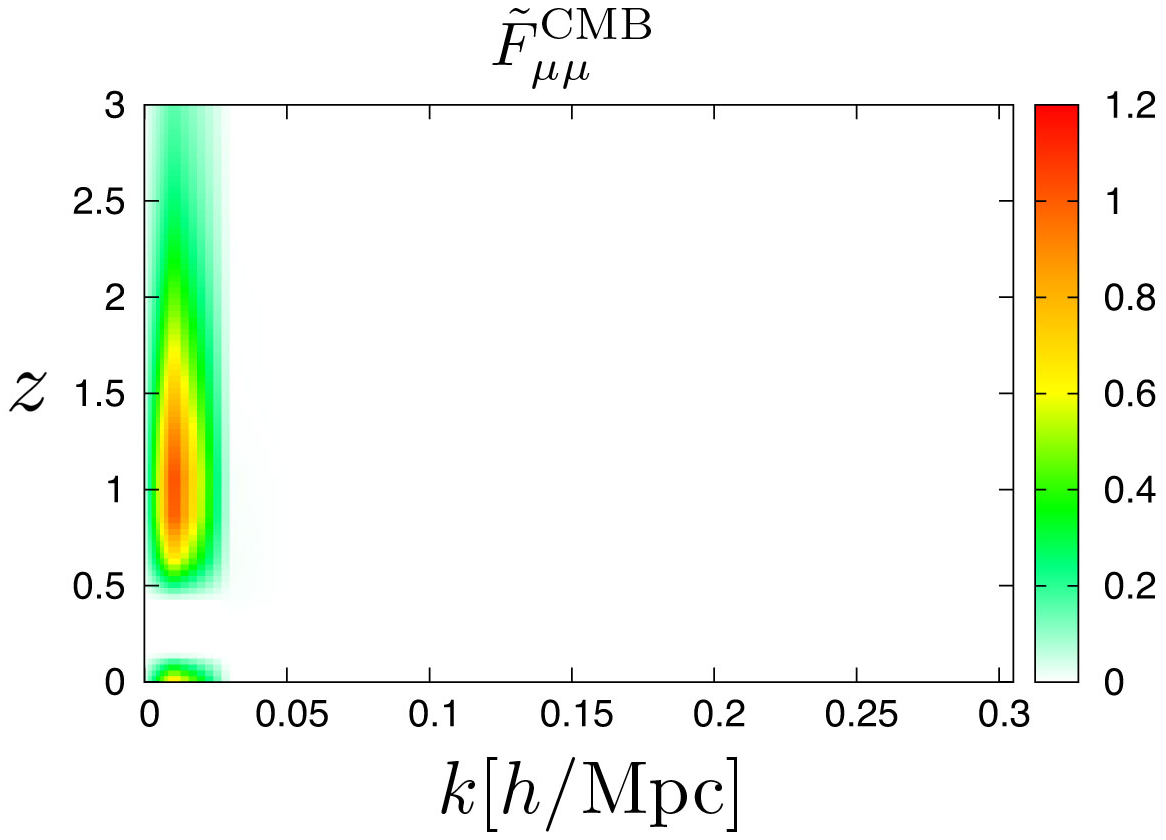}
	\hfill
	\includegraphics[width=.45\textwidth]{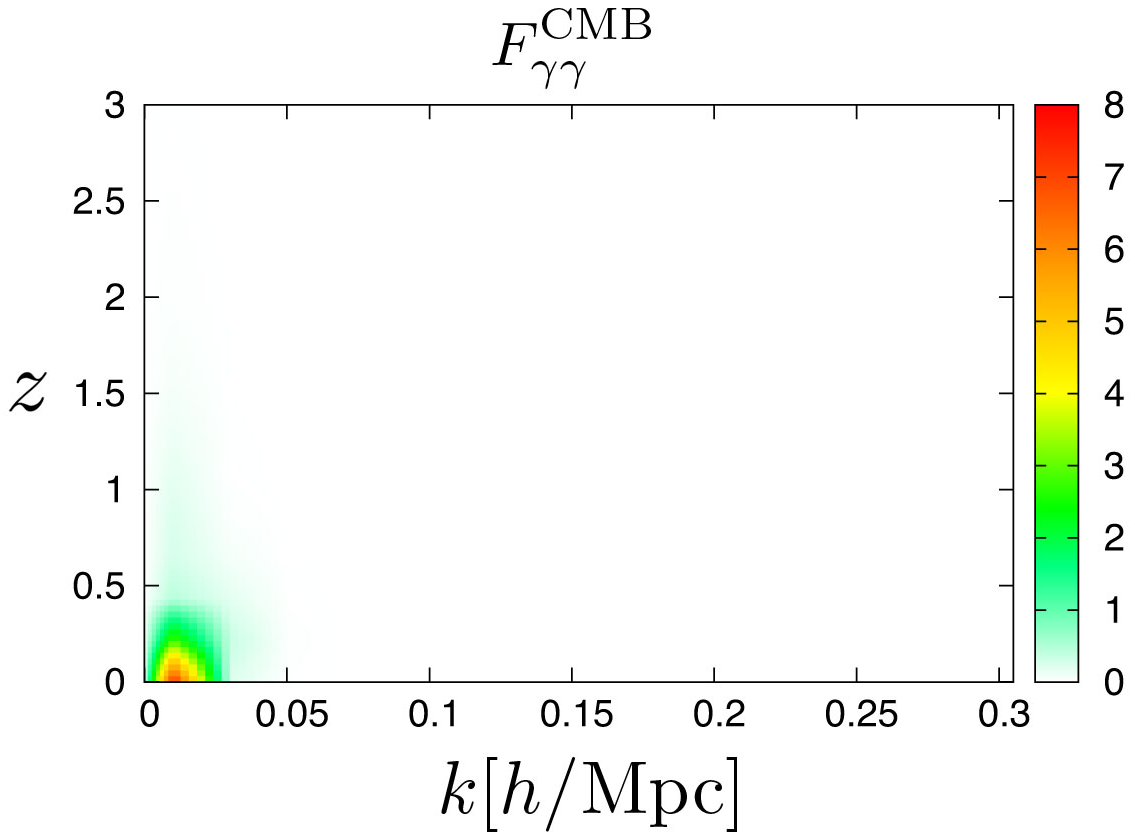}
	\newline
	\centering
	\includegraphics[width=.45\textwidth]{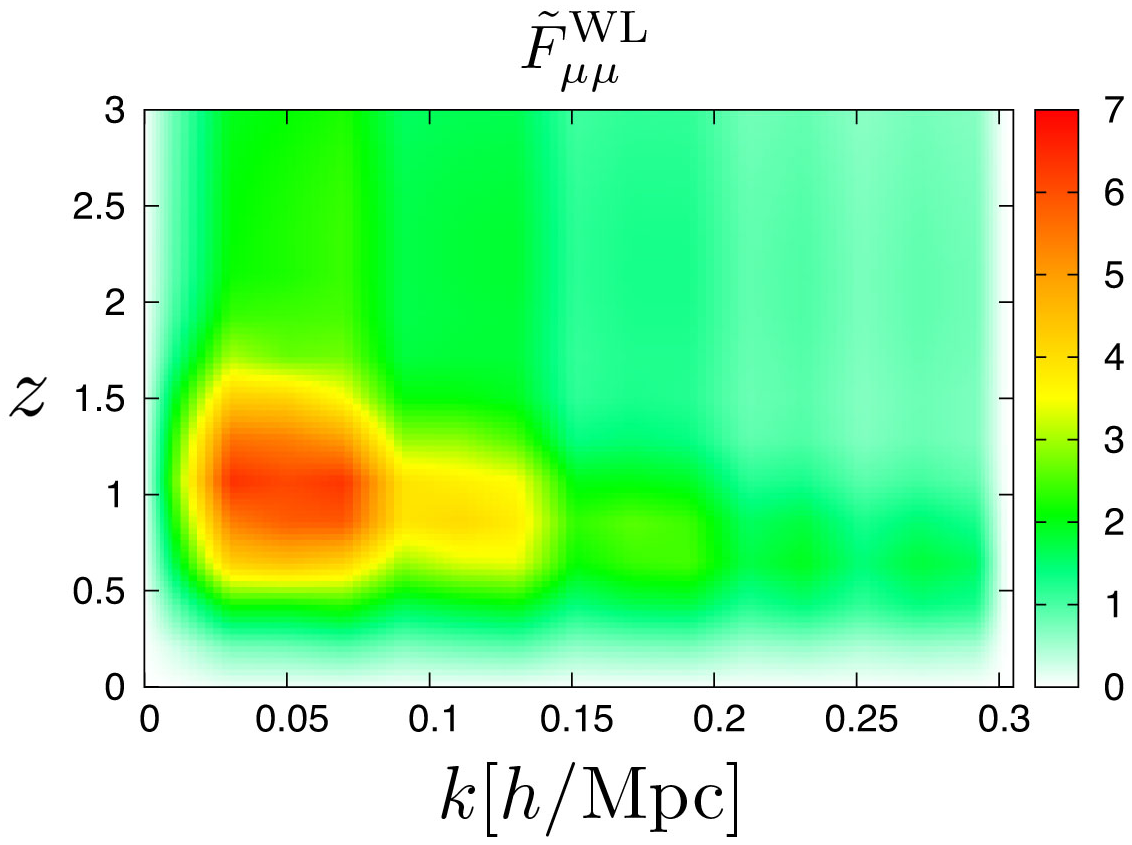}
	\hfill
	\includegraphics[width=.45\textwidth]{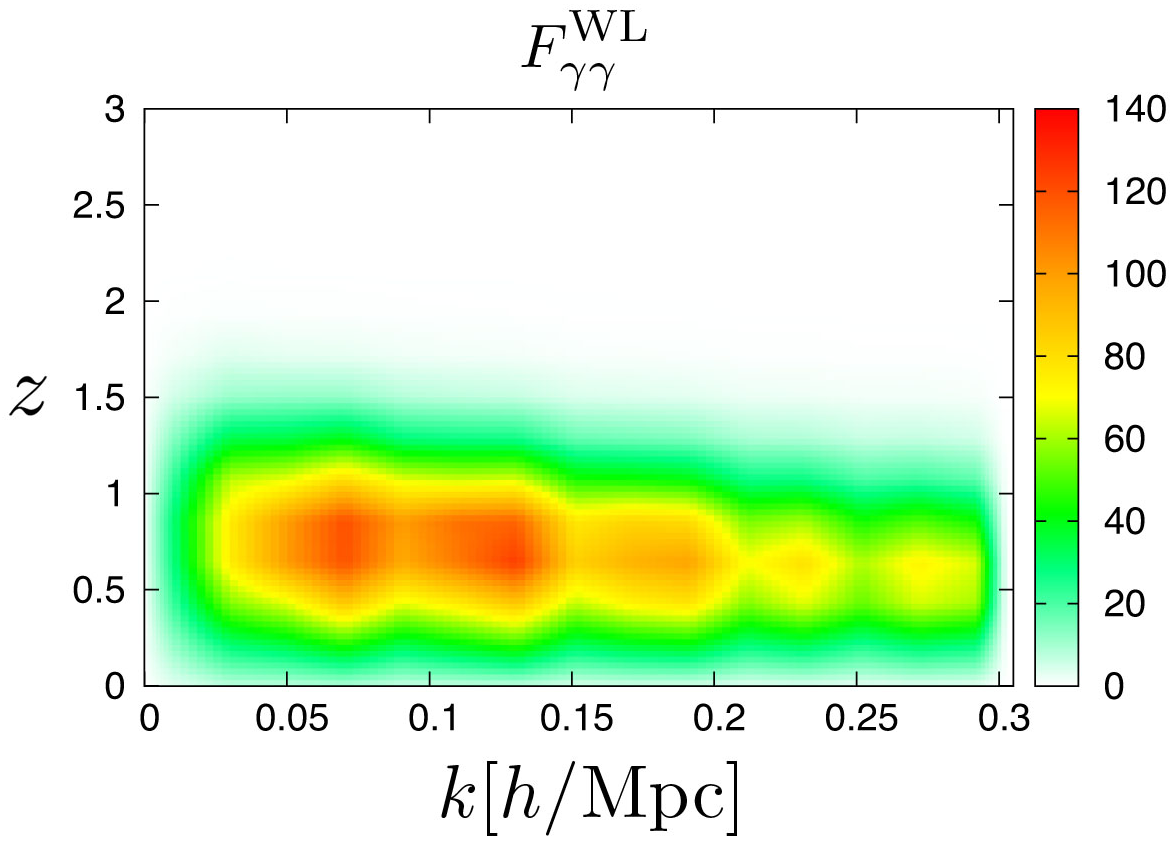}
	\newline
	\centering
	\includegraphics[width=.45\textwidth]{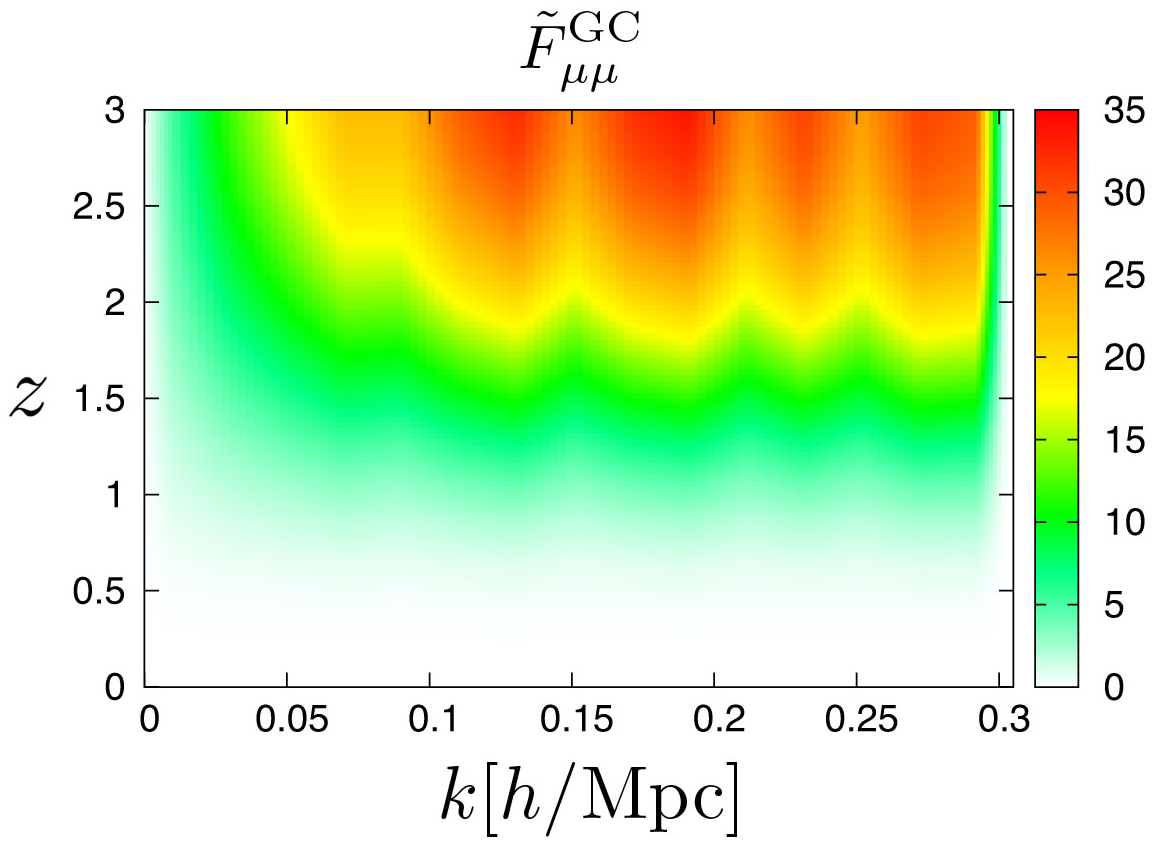}
	\hfill
	\includegraphics[width=.45\textwidth]{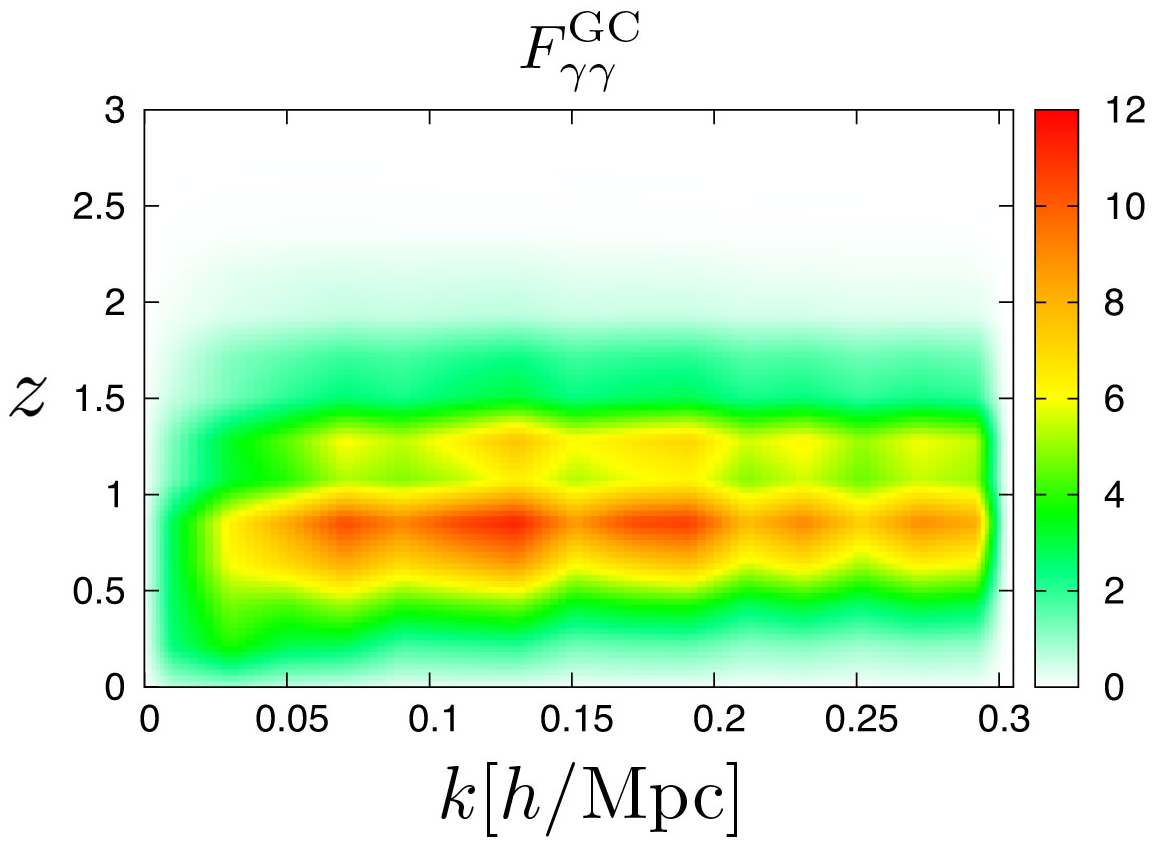}
	\newline
  \caption{The contributions of CMB, WL and GC (in the order from top to bottom) auto correlations to
    $\tilde{F}_{\mu\mu}^{\rm 2D}$ (left panels) and $F_{\gamma\gamma}^{\rm 2D}$ (right panels).}
  \label{fig:7}
\end{figure}
Assuming that $\mu$ and $\gamma$ are maximally correlated, i.e.,
$F^{\rm 2D}_{\mu\gamma}=\sqrt{F^{\rm 2D}_{\mu\mu}F^{\rm 2D}_{\gamma\gamma}} \sim 2 F^{\rm 2D}_{\gamma \gamma}$, we find that the total Fisher matrix from 2D measurements can be approximated as
\begin{align}
  {\bf F}=\left(\begin{array}{cc}F^{\rm 2D}_{\gamma\gamma} & F^{\rm 2D}_{\gamma\mu} \\F^{\rm 2D}_{\mu\gamma} &F^{\rm 2D}_{\mu\mu}  \end{array}\right)=\left(\begin{array}{cc}F^{\rm 2D}_{\gamma\gamma}  &2F^{\rm 2D}_{\gamma\gamma}   \\ 2F^{\rm 2D}_{\gamma\gamma} &4F^{\rm 2D}_{\gamma\gamma}+\tilde{F}^{\rm 2D}_{\mu\mu}  \end{array}\right) , \label{eq:5-3}
\end{align}
where $\tilde{F}^{\rm 2D}_{\mu\mu} \equiv F^{\rm 2D}_{\mu\mu} - 4F^{\rm 2D}_{\gamma\gamma}$.
As the value of $\tilde{F}^{\rm 2D}_{\mu\mu}$ is smaller, the degeneracy between $\mu$ and $\gamma$ is larger. The determinant of the Fisher matrix becomes $\det{{\bf F}}=F^{\rm 2D}_{\gamma\gamma}\tilde{F}^{\rm 2D}_{\mu\mu}$, and the
covariance matrix is given by
\begin{align}
	{\bf C}={\bf F}^{-1}=\left(\begin{array}{cc}4(\tilde{F}_{\mu\mu}^{\rm 2D})^{-1}+(F_{\gamma\gamma}^{\rm 2D})^{-1}  &-2(\tilde{F}_{\mu\mu}^{\rm 2D})^{-1}   \\ -2(\tilde{F}_{\mu\mu}^{\rm 2D})^{-1} &(\tilde{F}_{\mu\mu}^{\rm 2D})^{-1}  \end{array}\right). \label{eq:5-4}
\end{align}

The variances of $\mu$ and $\gamma$ are respectively given in terms of $\tilde{F}_{\mu\mu}^{\rm 2D}$ and $F_{\gamma\gamma}^{\rm 2D}$ as
\begin{align}
	C_{\gamma\gamma}&=4(\tilde{F}_{\mu\mu}^{\rm 2D})^{-1}+(F_{\gamma\gamma}^{\rm 2D})^{-1},\label{eq:5-5a}\\
	C_{\mu\mu}&=(\tilde{F}_{\mu\mu}^{\rm 2D})^{-1}, \label{eq:5-5b}
\end{align}
and, from eq.~(\ref{eq:15}), the variance of $\Sigma$ is given by
\begin{align}
	C_{\Sigma\Sigma}=(4F_{\gamma\gamma}^{\rm 2D})^{-1}.\label{eq:5-6}
\end{align}
From eqs.~(\ref{eq:5-5b}) and (\ref{eq:5-6}), $\tilde{F}_{\mu\mu}^{\rm 2D}$ and $F_{\gamma\gamma}^{\rm 2D}$ represent the inverse of covariance of $\mu$ and $\Sigma$ respectively.  From eq.~(\ref{eq:5-5a}), we find that both the values of $\tilde{F}_{\mu\mu}^{\rm 2D}$ and $F_{\gamma\gamma}^{\rm 2D}$ have to be larger than unity to constrain $\gamma$ well.

However, it is difficult to obtain large $\tilde{F}_{\mu\mu}^{\rm 2D}$ and $F_{\gamma\gamma}^{\rm 2D}$ simultaneously at any redshifts from 2D measurements only. Figure~\ref{fig:5-1} shows the values of $\tilde{F}_{\mu\mu}^{\rm 2D}$ and $F_{\gamma\gamma}^{\rm 2D}$ at each $k$ and $z$.  The value of $\tilde{F}_{\mu\mu}^{\rm 2D}$ is larger at higher redshifts because a change of $\mu$ at high $z$ affects the growth of structure at all lower $z$. On the other hand,
$F_{\gamma\gamma}^{\rm 2D}$ has large values at low redshifts ($z \simlt 1$) because WL is only sensitive to the change of $\gamma$ between the source galaxies and the observer. Such differences in redshift ranges where $\mu$ and $\gamma$ can be sensitively probed make it difficult to reduce the degeneracy between $\mu$ and $\gamma$ at all redshifts. Furthermore, the constraints on $\mu$ from 2D measurements are much weaker ($\tilde{F}_{\mu\mu}^{\rm 2D}$ is smaller) compared to the constraints on $\Sigma$, and thus the constraints on $\mu$ and $\gamma$ become weak.

Figure~\ref{fig:7} shows the contribution of each 2D measurement, i.e., CMB, WL, and GC to $\tilde{F}_{\mu\mu}^{\rm 2D}$ (left) and $F_{\gamma\gamma}^{\rm 2D}$ (right). Cross-correlations among different measurements are not included in this figure. We can see that the main contribution to $\tilde{F}_{\mu\mu}^{\rm 2D}$ comes from GC, while the main contribution to $F_{\gamma\gamma}^{\rm 2D}$ comes from WL, which is as expected. The ISW effect in CMB affects $F_{\gamma\gamma}^{\rm CMB}$ at small $k$ and low $z$. From figure~\ref{fig:7}, we find the contribution of ISW is smaller than other contributions, because the ISW effect appears on large scales where the number of modes, given by $2\ell+1$, is smaller. Moreover we can assume that $\tilde{F}^{\rm CMB}_{\mu\mu}$ is zero. Thus, CMB is not directly constraining the modified gravity parameters.
However, it helps to reduce the degeneracies between cosmological parameters and the MG parameters, thus tightening the constraints on MG parameters after marginalizing over cosmological parameters \cite{Hojjati:2012ci}. The contribution of WL to $\tilde{F}_{\mu\mu}^{\rm WL}$ is non-zero because the number distribution of source galaxies is changed by $\mu$ at high-$k$. $F^{\rm GC}_{\gamma\gamma}$ in figure~\ref{fig:7} comes from the magnification bias effects. The reason that $F^{\rm 2D}_{\gamma\gamma}\gg F^{\rm CMB}_{\gamma\gamma}+F^{\rm WL}_{\gamma\gamma}+F^{\rm GC}_{\gamma\gamma}$ is that the cross correlations, especially ${\rm WL}\times {\rm GC}$ (galaxy-galaxy lensing) are powerful tools for deriving information from weak lensing.

\begin{figure}[tbp]
  \centering
    \includegraphics[width=.45\textwidth]{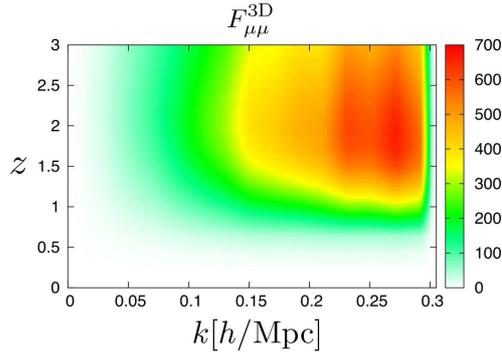}
    \caption{Fisher matrix of $\mu$ from the 3D galaxy power
    spectra $F_{\mu\mu}^{3D}$ (eq.~(\ref{eq:2-13})).}
  \label{fig:5-2}
\end{figure}

Next we see how the situation changes by adding in the 3D information. We denote the Fisher information of $\mu$ and $\gamma$ from 3D as
$F_{\mu\mu}^{\rm 3D}$ and $F_{\gamma\gamma}^{\rm 3D}$, respectively, and we assume $F_{\gamma\gamma}^{\rm 3D}=0$. The determinant of the Fisher matrix and the variances of $\mu$ and $\gamma$ are given by
\begin{align}
	\det{\bf F}&=(\tilde{F}_{\mu\mu}^{\rm 2D}+F^{\rm 3D}_{\mu\mu})F_{\gamma\gamma}^{\rm 2D}, \label{eq:5-7}\\
	C_{\gamma\gamma}&=4(\tilde{F}_{\mu\mu}^{\rm 2D}+F^{\rm 3D}_{\mu\mu})^{-1}+(F_{\gamma\gamma}^{\rm 2D})^{-1},\label{eq:5-8}\\
	C_{\mu\mu}&=(\tilde{F}_{\mu\mu}^{\rm 2D}+F^{\rm 3D}_{\mu\mu})^{-1}. \label{eq:5-9}
\end{align}
One can see that $F^{3D}_{\mu\mu}$ plays the same role as $\tilde{F}^{\rm 2D}_{\mu\mu}$ in reducing the degeneracy between $\mu$ and $\gamma$ because the correlation coefficient becomes
\begin{align}
	\frac{C_{\gamma\mu}}{\sqrt{C_{\gamma\gamma}C_{\mu\mu}}}&=-\left(1+\frac{\tilde{F}_{\mu\mu}^{\rm 2D}}{4F_{\gamma\gamma}^{\rm 2D}}\right)^{-1/2}\ {\rm (2D\ only)},\notag\\
	&= -\left(1+\frac{\tilde{F}_{\mu\mu}^{\rm 2D}+F^{\rm 3D}_{\mu\mu}}{4F_{\gamma\gamma}^{\rm 2D}}\right)^{-1/2}\ ({\rm 2D+3D}).\label{eq:5-10}
\end{align}

\begin{figure}[tbp]
\centering
	\includegraphics[width=.45\textwidth]{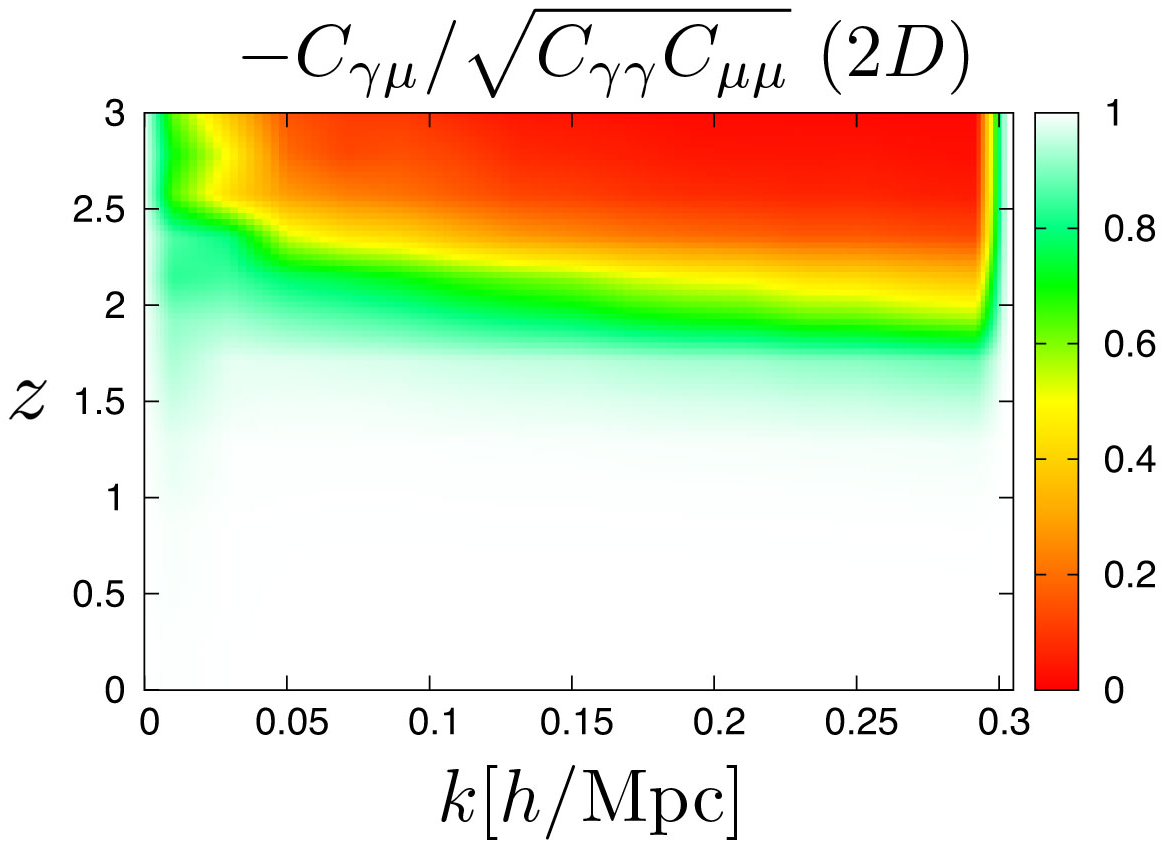}
\hfill
\centering
	\includegraphics[width=.45\textwidth]{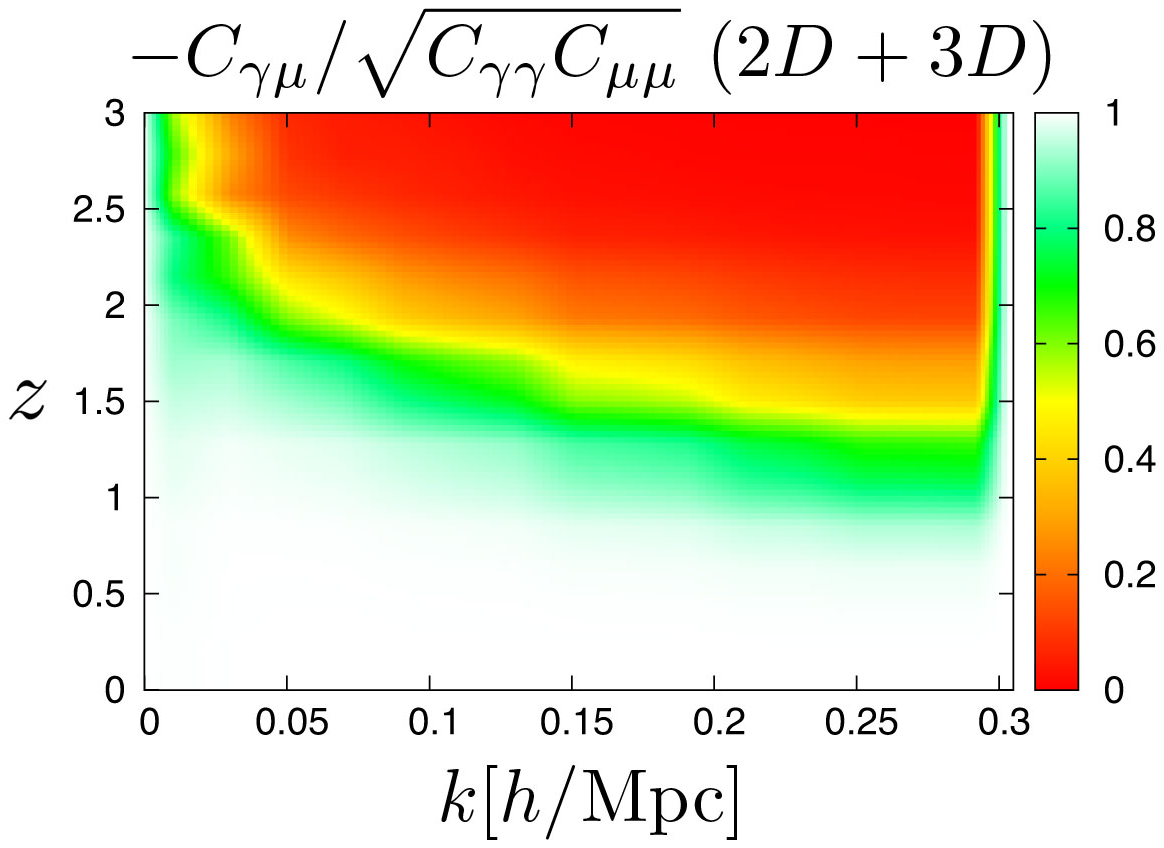}
	\caption{The correlation coefficient eq.~(\ref{eq:5-10}) in $(k,z)$ space by using only 2D only (left) and 2D+3D (right)}
	\label{fig:cc}
\end{figure}

Figure~\ref{fig:5-2} shows that $F_{\mu\mu}^{\rm 3D}$ is much larger than $\tilde{F}^{\rm 2D}_{\mu\mu}$ in the left panel of figure~\ref{fig:5-1}.
Therefore, the 3D information of the galaxy power spectra significantly improves the constraints on the parameters of $\mu$ and $\gamma$ and reduce the degeneracy between $\mu$ and $\gamma$.
Figure~\ref{fig:cc} shows the correlation coefficient between $\mu$ and $\gamma$ with and without the 3D information. The correlation coefficient is almost unity at $z<2$, meaning that $\mu$ and $\gamma$ have perfect degeneracy. However, the parameter space for the maximal degeneracy gets narrowed down to $z\lesssim1.5$ when the 3D information is added in. The reason for this degeneracy at $z\lesssim1.5$ is that the signal of the weak lensing dominates here.
However, we find that the degeneracy between $\mu$ and $\gamma$ becomes weaker by adding the 3D information.
Quantitatively, it results in a factor between $3$ and $10$ reduction of the errors over the redshift range that galaxy spectroscopic survey covers.

Note that, in principle the 3D information is not independent of GC, thus the correlation needs to be taken into account when combining them. But we ignore the correlation here since $F^{\rm 3D}_{\mu\mu}\gg\tilde{F}_{\mu\mu}^{\rm 2D}$ as seen from figure~\ref{fig:5-2} and the left panel of figure~\ref{fig:5-1}.

\section{Summary and Conclusions}\label{sec:6}
We performed a principal component analysis (PCA) to forecast the ability of future observations to measure departures from GR in Poisson, anisotropy and lensing potential equations parametrized by $\mu(k,z)$, $\gamma(k,z)$ and $\Sigma(k,z)$, respectively, in a model-independent way. We found that the galaxy power spectra in the redshift-space is a powerful tool for constraining $\mu$ and $\gamma$.
Combining galaxy redshift surveys like BOSS and Euclid surveys with weak lensing tomography decreases the error on principal component modes by a factor between $3$ and $10$, and increases the number of informative modes approximately by a factor of $3$ for $\mu$ and  by $7$ for $\gamma$. Such a gain in constraining MG parameters mainly comes from the fact that redshift-space distortion measurements from galaxy redshift surveys are a powerful tool to constrain $\mu$ and reduce the degeneracy between $\mu$ and $\gamma$, while lensing only constrains their combinations. Further, as shown in table~\ref{ta:2}, we find that the constraints on parameters of the modified gravity remain strong even if we take into account the uncertainty of cosmological parameters, because CMB measurements
strongly constrain the cosmological parameters.

Our analysis neglects various systematic errors. For lensing, these include possible shifts of the centroids and the dispersion of photo $z$-bins photometric redshift error, and the shape measurement error (see appendix in \cite{Hojjatietal:12} for details).  We also adopt a linear Kaiser formula to describe the redshift-space galaxy power spectra. More accurate formulae using $P_{\delta\theta}$ and velocity auto spectrum $P_{\theta\theta}$ are proposed by \cite{Scoccimarro:04,Matsubara:08,Taruyaetal:10}.  We neglect various non-linear effects on the galaxy clustering in redshift
space. Nonlinearity in galaxy biasing should be important at small scales and increases the uncertainty in measuring $\mu$. Nonlinear
redshift distortion effect, i.e., Finger-of-God effects, generates systematic uncertainties in the measurements of RSD \cite{HikageYamamoto:13}. Cross-correlation measurements with WL decreases the uncertainty of FoG effect by measuring the off-centering of satellite galaxies \cite{Hikageetal:12a}.  Such improvements in the theoretical modeling are left for future work.

\acknowledgments{
SA and CH are supported in part by Grant-in-Aid for Scientific
researcher of Japanese Ministry of Education, Culture, Sports, Science
and Technology (No. 24740160 for CH). KK is supported by STFC grant ST/H002774/1 and
ST/K0090X/1, the European Research Council and the Leverhulme trust.
KK thanks the Kobayashi-Maskawa Institute for the Origin of
Particles and the Universe for its hospitality during his visit when this work
was initiated. GBZ is supported by the {\it 1000 young talents} program in China, and by the University of Portsmouth. AH is supported by World Class University grant R32-2009-000-10130-0 through the National Research Foundation, Ministry of Education, Science and Technology of Korea. AH thanks Eric Linder for useful discussions. LP is supported by an NSERC Discovery grant.
}

\appendix
\section{Combined modes}
In the main text, we showed the eigenmodes obtained by diagonalizing
independently the covariance matrix for $\mu$, $\gamma$ or $\Sigma$.
We can instead obtain the eigenmodes in the whole $(\mu,\gamma)$ or
$(\mu,\Sigma)$ parameter space. In other words, we can diagonalize the
$2mn\times2mn$ covariance matrix which contains $\mu$,
($\gamma$ or $\Sigma$) and cross-covariance of the two parameters. We call
the eigenmodes obtained in this way the $(\mu,\gamma)$ or
$(\mu,\Sigma)$ combined modes. We also call the part of the combined
eigenmode associated with $\mu$ ($\gamma$ or $\Sigma$) the $\mu$
($\gamma$ or $\Sigma$) surface.

\begin{figure}[tpb]
  \centering
    \includegraphics[width=.65\textwidth]{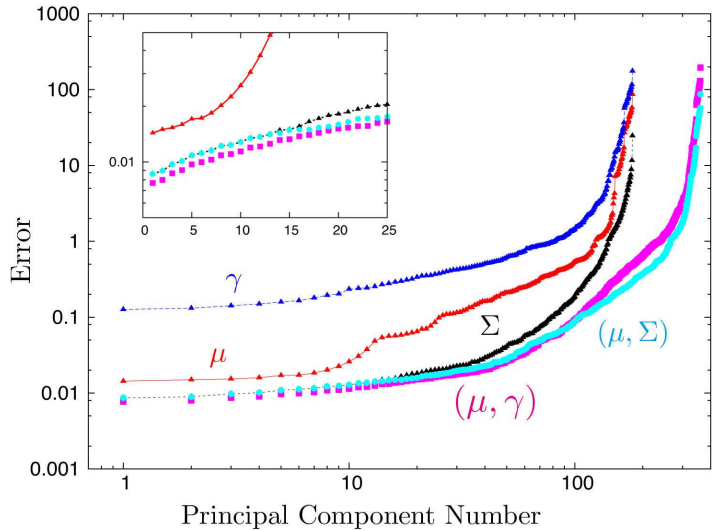}

  \caption{The error of the eigenmodes calculated from 2D and 3D. The
    triangle dots with the lines are the same as figure~\ref{fig:2}. The
    magenta square dots and the cyan circle dots are the error on the
    $(\mu,\gamma)$ and $(\mu,\Sigma)$ combined modes respectively. The
    mini-figure show the results enlarged where the principle
    component number is smaller. }
  \label{fig:A1}
\end{figure}

\begin{figure}[tbp]
\centering
	\includegraphics[width=.45\textwidth]{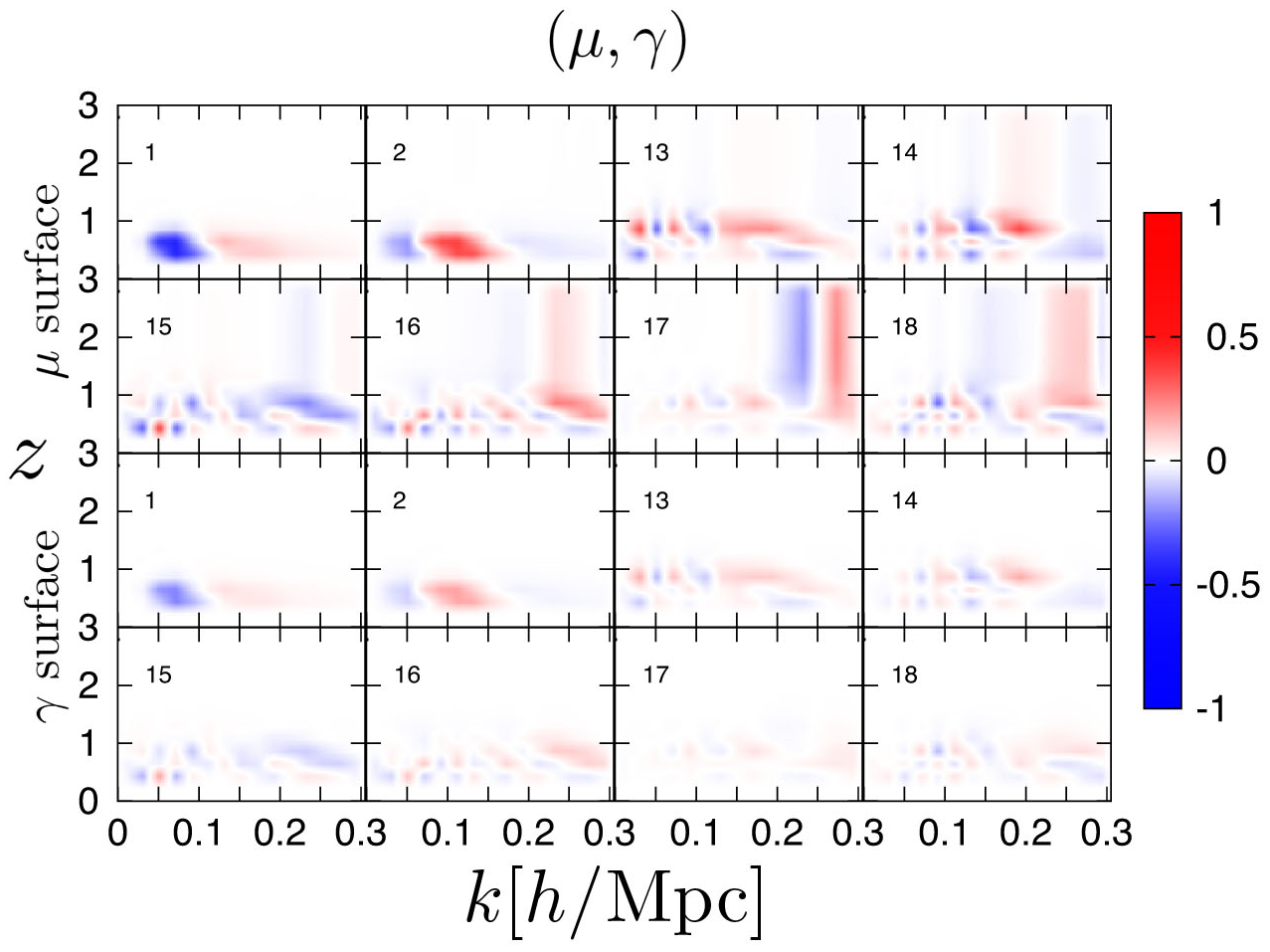}
\hfill
\centering
	\includegraphics[width=.45\textwidth]{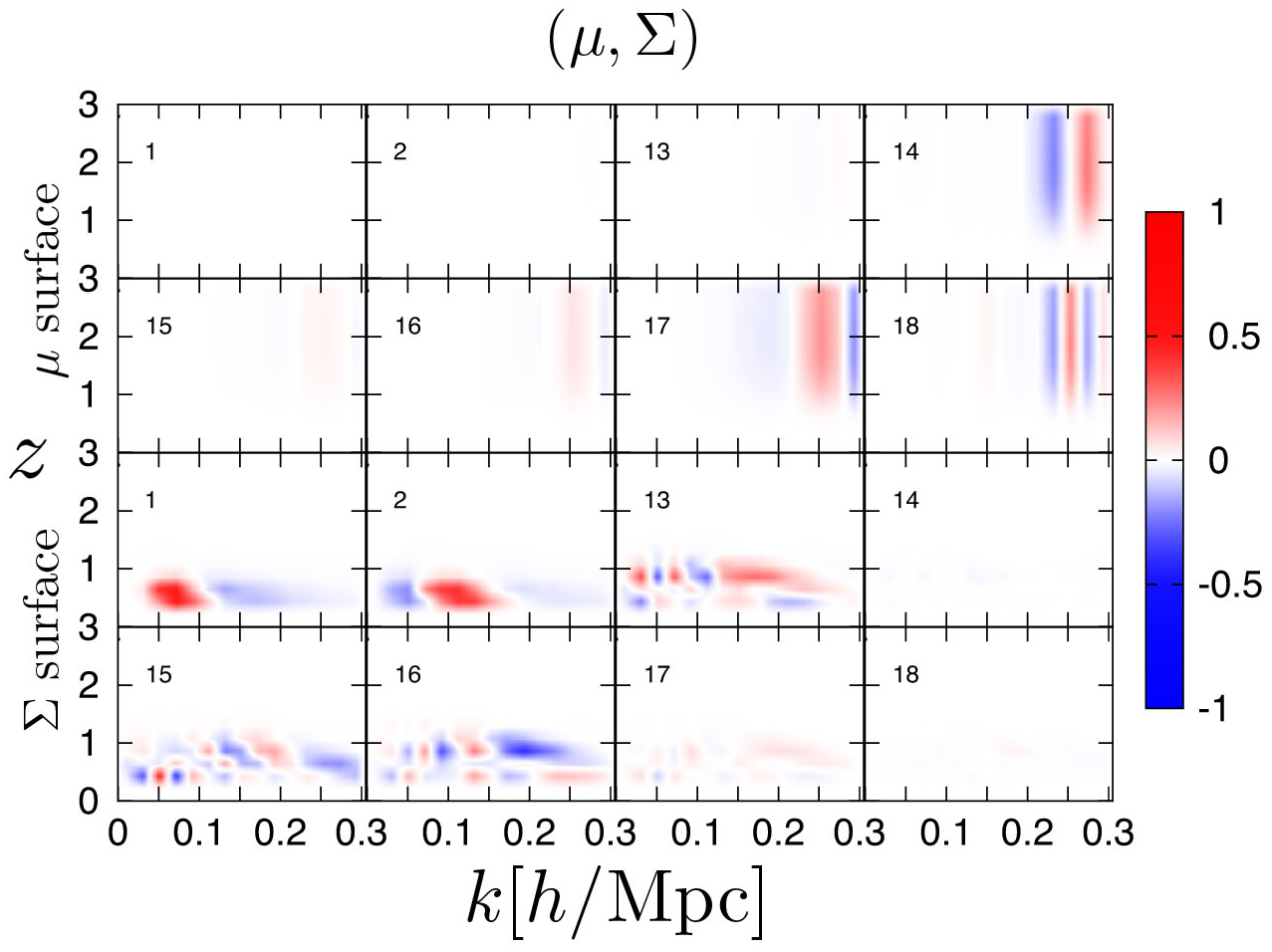}
  \caption{The parameter-surfaces of the combined modes. The left
    panel show the eigenmodes of the $(\mu,\gamma)$ combined modes and
    the right panel show the eigenmodes of the $(\mu,\Sigma)$ combined
    mode.}
  \label{fig:A2}
\end{figure}

Adding the information of the galaxy power spectra in the
redshift-space increases the numbers of eigenmodes of the
$(\mu,\gamma)$ and $(\mu,\Sigma)$ combined modes with $\sigma<1$ from
141 to 237 and from 152 to 265, respectively. In the following, we
show the result after adding the 3D information. Figure~\ref{fig:A1}
shows the error of the $(\mu,\gamma)$ or $(\mu,\Sigma)$ combined
modes, and figure~\ref{fig:A2} shows the combined modes.

From figure~\ref{fig:A2}, we find that the $\mu$-surfaces and the
$\Sigma$-surfaces of the $(\mu,\Sigma)$ combined modes are almost
independent. In other words, when one of the surfaces has nodes,
i.e. the regions that are constrained better, the other surface is
hardly constrained.  The reason is that $\mu$ and $\Sigma$ are
individually constrained from the galaxy distribution and the weak
lensing respectively as we expect from eqs.~(\ref{eq:5-6}) and (\ref{eq:5-9}). Moreover, the best-constrained eigenmode is the
eigenmode constraining $\Sigma$, because $\max{[4F^{\rm 2D}_{\gamma\gamma}(k.z)]}>\max{[\tilde{F}_{\mu\mu}^{\rm 2D}(k,z)+F^{\rm
      3D}_{\mu\mu}(k,z)]}$. The eigenmode that has a distinct node on
the $\mu$-surface appears in the 14th $(\mu,\Sigma)$ combined
eigenmode because from figure~\ref{fig:A1} the error on the
best-constrained eigenmode of $\mu$ is larger than the error on the
13th eigenmode of $\Sigma$ and smaller than the error on the 14th
eigenmodes of $\Sigma$.

On the other hand, the $\mu$-surfaces and $\gamma$-surfaces of the
$(\mu,\gamma)$ combined modes have the same shapes and they have
similar shapes to the eigenmodes of $\Sigma$ up to the 16th
eigenmodes. This is because $\Sigma$ is expressed as a combination of
$\mu$ and $\gamma$, and the constraints from the weak lensing is the
strongest.  The error on the best-constrained $(\mu,\gamma)$ combined
mode is smaller than that of the $(\mu,\Sigma)$ combined modes
because there is a degree of freedom in dividing the deviation of weak
lensing measurements from $\Lambda$CDM into $\mu$ and $\gamma$, and
$\mu$ at low-z, where we gain the most information from weak lensing,
is constrained only weakly by the galaxy distribution.

Moreover the shape of $(\mu,\gamma)$ combined eigenmodes is a mixture
of the $\mu$ and $\Sigma$ eigenmodes. For example, the elongated
feature in $z$-direction at $k=0.2-0.3$ clearly seen in the 16 or 17th
mode comes from the first eigenmode of $\mu$ (see the top-right panel
in figure~\ref{fig:3}). This can be understood from the minipanel of
figure~\ref{fig:A1} because the error of the best-constrained mode of
$\mu$ is nearly equal to the error of the 17th $(\mu,\gamma)$ combined
mode.

\bibliographystyle{JHEP}
\bibliography{jcapexample2}

\providecommand{\href}[2]{#2}\begingroup\raggedright\begin{thebibliography}{10}

\bibitem{A.G.Riessetal:98}
A.~G. {Riess}, A.~V. {Filippenko}, P.~{Challis}, A.~{Clocchiatti},
  A.~{Diercks}, P.~M. {Garnavich}, R.~L. {Gilliland}, C.~J. {Hogan}, S.~{Jha},
  R.~P. {Kirshner}, B.~{Leibundgut}, M.~M. {Phillips}, D.~{Reiss}, B.~P.
  {Schmidt}, R.~A. {Schommer}, R.~C. {Smith}, J.~{Spyromilio}, C.~{Stubbs},
  N.~B. {Suntzeff}, and J.~{Tonry}, {\it {Observational Evidence from
  Supernovae for an Accelerating Universe and a Cosmological Constant}},  {\em
  \aj} {\bf 116} (Sept., 1998) 1009--1038,
  [\href{http://xxx.lanl.gov/abs/astro-ph/9805201}{{\tt astro-ph/9805201}}].

\bibitem{S.Perlmutteretal:99}
S.~{Perlmutter}, G.~{Aldering}, G.~{Goldhaber}, R.~A. {Knop}, P.~{Nugent},
  P.~G. {Castro}, S.~{Deustua}, S.~{Fabbro}, A.~{Goobar}, D.~E. {Groom}, I.~M.
  {Hook}, A.~G. {Kim}, M.~Y. {Kim}, J.~C. {Lee}, N.~J. {Nunes}, R.~{Pain},
  C.~R. {Pennypacker}, R.~{Quimby}, C.~{Lidman}, R.~S. {Ellis}, M.~{Irwin},
  R.~G. {McMahon}, P.~{Ruiz-Lapuente}, N.~{Walton}, B.~{Schaefer}, B.~J.
  {Boyle}, A.~V. {Filippenko}, T.~{Matheson}, A.~S. {Fruchter}, N.~{Panagia},
  H.~J.~M. {Newberg}, W.~J. {Couch}, and {Supernova Cosmology Project}, {\it
  {Measurements of Omega and Lambda from 42 High-Redshift Supernovae}},  {\em
  \apj} {\bf 517} (June, 1999) 565--586,
  [\href{http://xxx.lanl.gov/abs/astro-ph/9812133}{{\tt astro-ph/9812133}}].

\bibitem{PLANCK}
{The Planck Collaboration}, {\it {The Scientific Programme of Planck}},  {\em
  ArXiv Astrophysics e-prints} (Apr., 2006)
  [\href{http://xxx.lanl.gov/abs/astro-ph/0604069}{{\tt astro-ph/0604069}}].

\bibitem{Sanchez:2012sg}
A.~G. {S{\'a}nchez}, C.~G. {Sc{\'o}ccola}, A.~J. {Ross}, W.~{Percival},
  M.~{Manera}, F.~{Montesano}, X.~{Mazzalay}, A.~J. {Cuesta}, D.~J.
  {Eisenstein}, E.~{Kazin}, C.~K. {McBride}, K.~{Mehta}, A.~D. {Montero-Dorta},
  N.~{Padmanabhan}, F.~{Prada}, J.~A. {Rubi{\~n}o-Mart{\'{\i}}n}, R.~{Tojeiro},
  X.~{Xu}, M.~V. {Maga{\~n}a}, E.~{Aubourg}, N.~A. {Bahcall}, S.~{Bailey},
  D.~{Bizyaev}, A.~S. {Bolton}, H.~{Brewington}, J.~{Brinkmann}, J.~R.
  {Brownstein}, J.~R. {Gott}, J.~C. {Hamilton}, S.~{Ho}, K.~{Honscheid},
  A.~{Labatie}, E.~{Malanushenko}, V.~{Malanushenko}, C.~{Maraston}, D.~{Muna},
  R.~C. {Nichol}, D.~{Oravetz}, K.~{Pan}, N.~P. {Ross}, N.~A. {Roe}, B.~A.
  {Reid}, D.~J. {Schlegel}, A.~{Shelden}, D.~P. {Schneider}, A.~{Simmons},
  R.~{Skibba}, S.~{Snedden}, D.~{Thomas}, J.~{Tinker}, D.~A. {Wake}, B.~A.
  {Weaver}, D.~H. {Weinberg}, M.~{White}, I.~{Zehavi}, and G.~{Zhao}, {\it {The
  clustering of galaxies in the SDSS-III Baryon Oscillation Spectroscopic
  Survey: cosmological implications of the large-scale two-point correlation
  function}},  {\em \mnras} {\bf 425} (Sept., 2012) 415--437,
  [\href{http://xxx.lanl.gov/abs/1203.6616}{{\tt arXiv:1203.6616}}].

\bibitem{Wetterich:88}
C.~{Wetterich}, {\it {Cosmology and the fate of dilatation symmetry}},  {\em
  Nuclear Physics B} {\bf 302} (June, 1988) 668--696.

\bibitem{Chibaetal:00}
T.~{Chiba}, T.~{Okabe}, and M.~{Yamaguchi}, {\it {Kinetically driven
  quintessence}},  {\em \prd} {\bf 62} (July, 2000) 023511,
  [\href{http://xxx.lanl.gov/abs/astro-ph/9912463}{{\tt astro-ph/9912463}}].

\bibitem{Bekenstein:04}
J.~D. {Bekenstein}, {\it {Relativistic gravitation theory for the modified
  Newtonian dynamics paradigm}},  {\em \prd} {\bf 70} (Oct., 2004) 083509,
  [\href{http://xxx.lanl.gov/abs/astro-ph/0403694}{{\tt astro-ph/0403694}}].

\bibitem{DGP}
G.~{Dvali}, G.~{Gabadadze}, and M.~{Porrati}, {\it {4D gravity on a brane in 5D
  Minkowski space}},  {\em Physics Letters B} {\bf 485} (July, 2000) 208--214,
  [\href{http://xxx.lanl.gov/abs/hep-th/0005016}{{\tt hep-th/0005016}}].

\bibitem{Milgrom:83}
M.~{Milgrom}, {\it {A modification of the Newtonian dynamics - Implications for
  galaxies}},  {\em \apj} {\bf 270} (July, 1983) 371--389.

\bibitem{Elingetal:04}
C.~{Eling}, T.~{Jacobson}, and D.~{Mattingly}, {\it {Einstein-Aether Theory}},
  {\em ArXiv General Relativity and Quantum Cosmology e-prints} (Sept., 2004)
  [\href{http://xxx.lanl.gov/abs/gr-qc/0410001}{{\tt gr-qc/0410001}}].

\bibitem{Capozziello:2003tk}
S.~{Capozziello}, S.~{Carloni}, and A.~{Troisi}, {\it {Quintessence without
  scalar fields}},  {\em ArXiv Astrophysics e-prints} (Mar., 2003)
  [\href{http://xxx.lanl.gov/abs/astro-ph/0303041}{{\tt astro-ph/0303041}}].

\bibitem{Carrolletal:04}
S.~M. {Carroll}, V.~{Duvvuri}, M.~{Trodden}, and M.~S. {Turner}, {\it {Is
  cosmic speed-up due to new gravitational physics?}},  {\em \prd} {\bf 70}
  (Aug., 2004) 043528, [\href{http://xxx.lanl.gov/abs/astro-ph/0306438}{{\tt
  astro-ph/0306438}}].

\bibitem{Nicolisetal:09}
A.~{Nicolis}, R.~{Rattazzi}, and E.~{Trincherini}, {\it {Galileon as a local
  modification of gravity}},  {\em \prd} {\bf 79} (Mar., 2009) 064036,
  [\href{http://xxx.lanl.gov/abs/0811.2197}{{\tt arXiv:0811.2197}}].

\bibitem{Clifton:2011}
T.~{Clifton}, P.~G. {Ferreira}, A.~{Padilla}, and C.~{Skordis}, {\it {Modified
  gravity and cosmology}},  {\em \physrep} {\bf 513} (Mar., 2012) 1--189,
  [\href{http://xxx.lanl.gov/abs/1106.2476}{{\tt arXiv:1106.2476}}].

\bibitem{Oyaizuetal:08}
H.~{Oyaizu}, M.~{Lima}, and W.~{Hu}, {\it {Nonlinear evolution of f(R)
  cosmologies. II. Power spectrum}},  {\em \prd} {\bf 78} (Dec., 2008) 123524,
  [\href{http://xxx.lanl.gov/abs/0807.2462}{{\tt arXiv:0807.2462}}].

\bibitem{Zhaoetal:11}
G.-B. {Zhao}, B.~{Li}, and K.~{Koyama}, {\it {N-body simulations for f(R)
  gravity using a self-adaptive particle-mesh code}},  {\em \prd} {\bf 83}
  (Feb., 2011) 044007, [\href{http://xxx.lanl.gov/abs/1011.1257}{{\tt
  arXiv:1011.1257}}].

\bibitem{Li:2011vk}
B.~{Li}, G.-B. {Zhao}, R.~{Teyssier}, and K.~{Koyama}, {\it {ECOSMOG: an
  Efficient COde for Simulating MOdified Gravity}},  {\em \jcap} {\bf 1} (Jan.,
  2012) 51, [\href{http://xxx.lanl.gov/abs/1110.1379}{{\tt arXiv:1110.1379}}].

\bibitem{Puchweinetal:13}
E.~{Puchwein}, M.~{Baldi}, and V.~{Springel}, {\it {Modified Gravity-GADGET: A
  new code for cosmological hydrodynamical simulations of modified gravity
  models}},  {\em ArXiv e-prints} (May, 2013)
  [\href{http://xxx.lanl.gov/abs/1305.2418}{{\tt arXiv:1305.2418}}].

\bibitem{DES}
``\url{http://www.darkenergysurvey.org/}.''

\bibitem{BOSS2}
``\url{http://www.sdss3.org/surveys/boss.php}.''

\bibitem{LSST}
``\url{http://www.lsst.org/}.''

\bibitem{SUMIRE}
``\url{http://sumire.ipmu.jp/en/}.''

\bibitem{BigBOSS}
D.~J. {Schlegel}, C.~{Bebek}, H.~{Heetderks}, S.~{Ho}, M.~{Lampton}, M.~{Levi},
  N.~{Mostek}, N.~{Padmanabhan}, S.~{Perlmutter}, N.~{Roe}, M.~{Sholl},
  G.~{Smoot}, M.~{White}, A.~{Dey}, T.~{Abraham}, B.~{Jannuzi}, D.~{Joyce},
  M.~{Liang}, M.~{Merrill}, K.~{Olsen}, and S.~{Salim}, {\it {BigBOSS: The
  Ground-Based Stage IV Dark Energy Experiment}},  {\em ArXiv e-prints} (Apr.,
  2009) [\href{http://xxx.lanl.gov/abs/0904.0468}{{\tt arXiv:0904.0468}}].

\bibitem{BigBOSS2}
``\url{http://bigboss.lbl.gov/}.''

\bibitem{HETDEX}
``\url{http://hetdex.org/}.''

\bibitem{euclid2}
``\url{http://www.euclid-ec.org}.''

\bibitem{G.B.Zhaoetal:09}
G.-B. {Zhao}, L.~{Pogosian}, A.~{Silvestri}, and J.~{Zylberberg}, {\it
  {Searching for modified growth patterns with tomographic surveys}},  {\em
  \prd} {\bf 79} (Apr., 2009) 083513,
  [\href{http://xxx.lanl.gov/abs/0809.3791}{{\tt arXiv:0809.3791}}].

\bibitem{Pogosianetal:10}
L.~{Pogosian}, A.~{Silvestri}, K.~{Koyama}, and G.-B. {Zhao}, {\it {How to
  optimally parametrize deviations from general relativity in the evolution of
  cosmological perturbations}},  {\em \prd} {\bf 81} (May, 2010) 104023,
  [\href{http://xxx.lanl.gov/abs/1002.2382}{{\tt arXiv:1002.2382}}].

\bibitem{Zhangetal:07}
P.~{Zhang}, M.~{Liguori}, R.~{Bean}, and S.~{Dodelson}, {\it {Probing Gravity
  at Cosmological Scales by Measurements which Test the Relationship between
  Gravitational Lensing and Matter Overdensity}},  {\em Physical Review
  Letters} {\bf 99} (Oct., 2007) 141302,
  [\href{http://xxx.lanl.gov/abs/0704.1932}{{\tt arXiv:0704.1932}}].

\bibitem{Amendolaetal:08}
L.~{Amendola}, M.~{Kunz}, and D.~{Sapone}, {\it {Measuring the dark side (with
  weak lensing)}},  {\em \jcap} {\bf 4} (Apr., 2008) 13,
  [\href{http://xxx.lanl.gov/abs/0704.2421}{{\tt arXiv:0704.2421}}].

\bibitem{Hu:2007pj}
W.~{Hu} and I.~{Sawicki}, {\it {Parametrized post-Friedmann framework for
  modified gravity}},  {\em \prd} {\bf 76} (Nov., 2007) 104043,
  [\href{http://xxx.lanl.gov/abs/0708.1190}{{\tt arXiv:0708.1190}}].

\bibitem{Baker:2011jy}
T.~{Baker}, P.~G. {Ferreira}, C.~{Skordis}, and J.~{Zuntz}, {\it {Towards a
  fully consistent parametrization of modified gravity}},  {\em \prd} {\bf 84}
  (Dec., 2011) 124018, [\href{http://xxx.lanl.gov/abs/1107.0491}{{\tt
  arXiv:1107.0491}}].

\bibitem{Zuntz:2011aq}
J.~{Zuntz}, T.~{Baker}, P.~G. {Ferreira}, and C.~{Skordis}, {\it {Ambiguous
  tests of general relativity on cosmological scales}},  {\em \jcap} {\bf 6}
  (June, 2012) 32, [\href{http://xxx.lanl.gov/abs/1110.3830}{{\tt
  arXiv:1110.3830}}].

\bibitem{Baker:2012zs}
T.~{Baker}, P.~G. {Ferreira}, and C.~{Skordis}, {\it {The parameterized
  post-Friedmann framework for theories of modified gravity: Concepts,
  formalism, and examples}},  {\em \prd} {\bf 87} (Jan., 2013) 024015,
  [\href{http://xxx.lanl.gov/abs/1209.2117}{{\tt arXiv:1209.2117}}].

\bibitem{Battye:2013er}
R.~A. {Battye} and J.~A. {Pearson}, {\it {Massive gravity, the elasticity of
  space-time and perturbations in the dark sector}},  {\em ArXiv e-prints}
  (Jan., 2013) [\href{http://xxx.lanl.gov/abs/1301.5042}{{\tt
  arXiv:1301.5042}}].

\bibitem{Zhaoetal:10}
G.-B. {Zhao}, T.~{Giannantonio}, L.~{Pogosian}, A.~{Silvestri}, D.~J. {Bacon},
  K.~{Koyama}, R.~C. {Nichol}, and Y.-S. {Song}, {\it {Probing modifications of
  general relativity using current cosmological observations}},  {\em \prd}
  {\bf 81} (May, 2010) 103510, [\href{http://xxx.lanl.gov/abs/1003.0001}{{\tt
  arXiv:1003.0001}}].

\bibitem{Samushiaetal:13}
L.~{Samushia}, B.~A. {Reid}, M.~{White}, W.~J. {Percival}, A.~J. {Cuesta},
  L.~{Lombriser}, M.~{Manera}, R.~C. {Nichol}, D.~P. {Schneider}, D.~{Bizyaev},
  H.~{Brewington}, E.~{Malanushenko}, V.~{Malanushenko}, D.~{Oravetz},
  K.~{Pan}, A.~{Simmons}, A.~{Shelden}, S.~{Snedden}, J.~L. {Tinker}, B.~A.
  {Weaver}, D.~G. {York}, and G.-B. {Zhao}, {\it {The clustering of galaxies in
  the SDSS-III DR9 Baryon Oscillation Spectroscopic Survey: testing deviations
  from {$\Lambda$} and general relativity using anisotropic clustering of
  galaxies}},  {\em \mnras} {\bf 429} (Feb., 2013) 1514--1528,
  [\href{http://xxx.lanl.gov/abs/1206.5309}{{\tt arXiv:1206.5309}}].

\bibitem{Simpsonetal:13}
F.~{Simpson}, C.~{Heymans}, D.~{Parkinson}, C.~{Blake}, M.~{Kilbinger},
  J.~{Benjamin}, T.~{Erben}, H.~{Hildebrandt}, H.~{Hoekstra}, T.~D. {Kitching},
  Y.~{Mellier}, L.~{Miller}, L.~{Van Waerbeke}, J.~{Coupon}, L.~{Fu},
  J.~{Harnois-D{\'e}raps}, M.~J. {Hudson}, K.~{Kuijken}, B.~{Rowe},
  T.~{Schrabback}, E.~{Semboloni}, S.~{Vafaei}, and M.~{Velander}, {\it
  {CFHTLenS: testing the laws of gravity with tomographic weak lensing and
  redshift-space distortions}},  {\em \mnras} {\bf 429} (Mar., 2013)
  2249--2263, [\href{http://xxx.lanl.gov/abs/1212.3339}{{\tt
  arXiv:1212.3339}}].

\bibitem{Zhao:2009fn}
G.-B. {Zhao}, L.~{Pogosian}, A.~{Silvestri}, and J.~{Zylberberg}, {\it
  {Cosmological Tests of General Relativity with Future Tomographic Surveys}},
  {\em Physical Review Letters} {\bf 103} (Dec., 2009) 241301,
  [\href{http://xxx.lanl.gov/abs/0905.1326}{{\tt arXiv:0905.1326}}].

\bibitem{Hojjatietal:12}
A.~{Hojjati}, G.-B. {Zhao}, L.~{Pogosian}, A.~{Silvestri}, R.~{Crittenden}, and
  K.~{Koyama}, {\it {Cosmological tests of general relativity: A principal
  component analysis}},  {\em \prd} {\bf 85} (Feb., 2012) 043508,
  [\href{http://xxx.lanl.gov/abs/1111.3960}{{\tt arXiv:1111.3960}}].

\bibitem{Hall:2012wd}
A.~{Hall}, C.~{Bonvin}, and A.~{Challinor}, {\it {Testing general relativity
  with 21-cm intensity mapping}},  {\em \prd} {\bf 87} (Mar., 2013) 064026,
  [\href{http://xxx.lanl.gov/abs/1212.0728}{{\tt arXiv:1212.0728}}].

\bibitem{Kaiser:87}
N.~{Kaiser}, {\it {Clustering in real space and in redshift space}},  {\em
  \mnras} {\bf 227} (July, 1987) 1--21.

\bibitem{Guzik:2009cm}
J.~{Guzik}, B.~{Jain}, and M.~{Takada}, {\it {Tests of gravity from imaging and
  spectroscopic surveys}},  {\em \prd} {\bf 81} (Jan., 2010) 023503,
  [\href{http://xxx.lanl.gov/abs/0906.2221}{{\tt arXiv:0906.2221}}].

\bibitem{Songetal:11}
Y.-S. {Song}, G.-B. {Zhao}, D.~{Bacon}, K.~{Koyama}, R.~C. {Nichol}, and
  L.~{Pogosian}, {\it {Complementarity of weak lensing and peculiar velocity
  measurements in testing general relativity}},  {\em \prd} {\bf 84} (Oct.,
  2011) 083523, [\href{http://xxx.lanl.gov/abs/1011.2106}{{\tt
  arXiv:1011.2106}}].

\bibitem{PLANCK16}
{Planck Collaboration}, P.~A.~R. {Ade}, N.~{Aghanim}, C.~{Armitage-Caplan},
  M.~{Arnaud}, M.~{Ashdown}, F.~{Atrio-Barandela}, J.~{Aumont},
  C.~{Baccigalupi}, A.~J. {Banday}, and et~al., {\it {Planck 2013 results. XVI.
  Cosmological parameters}},  {\em ArXiv e-prints} (Mar., 2013)
  [\href{http://xxx.lanl.gov/abs/1303.5076}{{\tt arXiv:1303.5076}}].

\bibitem{Euclid}
L.~{Amendola}, S.~{Appleby}, D.~{Bacon}, T.~{Baker}, M.~{Baldi}, N.~{Bartolo},
  A.~{Blanchard}, C.~{Bonvin}, S.~{Borgani}, E.~{Branchini}, C.~{Burrage},
  S.~{Camera}, C.~{Carbone}, L.~{Casarini}, M.~{Cropper}, C.~{deRham}, C.~{di
  Porto}, A.~{Ealet}, P.~G. {Ferreira}, F.~{Finelli}, J.~{Garcia-Bellido},
  T.~{Giannantonio}, L.~{Guzzo}, A.~{Heavens}, L.~{Heisenberg}, C.~{Heymans},
  H.~{Hoekstra}, L.~{Hollenstein}, R.~{Holmes}, O.~{Horst}, K.~{Jahnke}, T.~D.
  {Kitching}, T.~{Koivisto}, M.~{Kunz}, G.~{La Vacca}, M.~{March},
  E.~{Majerotto}, K.~{Markovic}, D.~{Marsh}, F.~{Marulli}, R.~{Massey},
  Y.~{Mellier}, D.~F. {Mota}, N.~{Nunes}, W.~{Percival}, V.~{Pettorino},
  C.~{Porciani}, C.~{Quercellini}, J.~{Read}, M.~{Rinaldi}, D.~{Sapone},
  R.~{Scaramella}, C.~{Skordis}, F.~{Simpson}, A.~{Taylor}, S.~{Thomas},
  R.~{Trotta}, L.~{Verde}, F.~{Vernizzi}, A.~{Vollmer}, Y.~{Wang}, J.~{Weller},
  and T.~{Zlosnik}, {\it {Cosmology and fundamental physics with the Euclid
  satellite}},  {\em ArXiv e-prints} (June, 2012)
  [\href{http://xxx.lanl.gov/abs/1206.1225}{{\tt arXiv:1206.1225}}].

\bibitem{Turneretal:84}
E.~L. {Turner}, J.~P. {Ostriker}, and J.~R. {Gott}, III, {\it {The statistics
  of gravitational lenses - The distributions of image angular separations and
  lens redshifts}},  {\em \apj} {\bf 284} (Sept., 1984) 1--22.

\bibitem{Villumsen:95}
J.~{Verner Villumsen}, {\it {Clustering of Faint Galaxies: Induced by Weak
  Gravitational Lensing}},  {\em ArXiv Astrophysics e-prints} (Dec., 1995)
  [\href{http://xxx.lanl.gov/abs/astro-ph/9512001}{{\tt astro-ph/9512001}}].

\bibitem{MGCAMB}
``\url{http://www.sfu.ca/~aha25/MGCAMB.html}.''

\bibitem{Hojjati:2011ix}
A.~{Hojjati}, L.~{Pogosian}, and G.-B. {Zhao}, {\it {Testing gravity with CAMB
  and CosmoMC}},  {\em \jcap} {\bf 8} (Aug., 2011) 5,
  [\href{http://xxx.lanl.gov/abs/1106.4543}{{\tt arXiv:1106.4543}}].

\bibitem{Tegmarketal:97}
M.~{Tegmark}, A.~N. {Taylor}, and A.~F. {Heavens}, {\it {Karhunen-Loeve
  Eigenvalue Problems in Cosmology: How Should We Tackle Large Data Sets?}},
  {\em \apj} {\bf 480} (May, 1997) 22,
  [\href{http://xxx.lanl.gov/abs/astro-ph/9603021}{{\tt astro-ph/9603021}}].

\bibitem{Kitching:2010ab}
T.~D. {Kitching} and A.~N. {Taylor}, {\it {On mitigation of the uncertainty in
  non-linear matter clustering for cosmic shear tomography}},  {\em \mnras}
  {\bf 416} (Sept., 2011) 1717--1722,
  [\href{http://xxx.lanl.gov/abs/1012.3479}{{\tt arXiv:1012.3479}}].

\bibitem{Tegmark:97}
M.~{Tegmark}, {\it {Measuring Cosmological Parameters with Galaxy Surveys}},
  {\em Physical Review Letters} {\bf 79} (Nov., 1997) 3806--3809,
  [\href{http://xxx.lanl.gov/abs/astro-ph/9706198}{{\tt astro-ph/9706198}}].

\bibitem{Gaztanagaetal:12}
E.~{Gazta{\~n}aga}, M.~{Eriksen}, M.~{Crocce}, F.~J. {Castander}, P.~{Fosalba},
  P.~{Marti}, R.~{Miquel}, and A.~{Cabr{\'e}}, {\it {Cross-correlation of
  spectroscopic and photometric galaxy surveys: cosmology from lensing and
  redshift distortions}},  {\em \mnras} {\bf 422} (June, 2012) 2904--2930,
  [\href{http://xxx.lanl.gov/abs/1109.4852}{{\tt arXiv:1109.4852}}].

\bibitem{BOSS}
K.~S. {Dawson}, D.~J. {Schlegel}, C.~P. {Ahn}, S.~F. {Anderson},
  {\'E}.~{Aubourg}, S.~{Bailey}, R.~H. {Barkhouser}, J.~E. {Bautista},
  A.~{Beifiori}, A.~A. {Berlind}, V.~{Bhardwaj}, D.~{Bizyaev}, C.~H. {Blake},
  M.~R. {Blanton}, M.~{Blomqvist}, A.~S. {Bolton}, A.~{Borde}, J.~{Bovy}, W.~N.
  {Brandt}, H.~{Brewington}, J.~{Brinkmann}, P.~J. {Brown}, J.~R. {Brownstein},
  K.~{Bundy}, N.~G. {Busca}, W.~{Carithers}, A.~R. {Carnero}, M.~A. {Carr},
  Y.~{Chen}, J.~{Comparat}, N.~{Connolly}, F.~{Cope}, R.~A.~C. {Croft}, A.~J.
  {Cuesta}, L.~N. {da Costa}, J.~R.~A. {Davenport}, T.~{Delubac}, R.~{de
  Putter}, S.~{Dhital}, A.~{Ealet}, G.~L. {Ebelke}, D.~J. {Eisenstein},
  S.~{Escoffier}, X.~{Fan}, N.~{Filiz Ak}, H.~{Finley}, A.~{Font-Ribera},
  R.~{G{\'e}nova-Santos}, J.~E. {Gunn}, H.~{Guo}, D.~{Haggard}, P.~B. {Hall},
  J.-C. {Hamilton}, B.~{Harris}, D.~W. {Harris}, S.~{Ho}, D.~W. {Hogg},
  D.~{Holder}, K.~{Honscheid}, J.~{Huehnerhoff}, B.~{Jordan}, W.~P. {Jordan},
  G.~{Kauffmann}, E.~A. {Kazin}, D.~{Kirkby}, M.~A. {Klaene}, J.-P. {Kneib},
  J.-M. {Le Goff}, K.-G. {Lee}, D.~C. {Long}, C.~P. {Loomis}, B.~{Lundgren},
  R.~H. {Lupton}, M.~A.~G. {Maia}, M.~{Makler}, E.~{Malanushenko},
  V.~{Malanushenko}, R.~{Mandelbaum}, M.~{Manera}, C.~{Maraston}, D.~{Margala},
  K.~L. {Masters}, C.~K. {McBride}, P.~{McDonald}, I.~D. {McGreer}, R.~G.
  {McMahon}, O.~{Mena}, J.~{Miralda-Escud{\'e}}, A.~D. {Montero-Dorta},
  F.~{Montesano}, D.~{Muna}, A.~D. {Myers}, T.~{Naugle}, R.~C. {Nichol},
  P.~{Noterdaeme}, S.~E. {Nuza}, M.~D. {Olmstead}, A.~{Oravetz}, D.~J.
  {Oravetz}, R.~{Owen}, N.~{Padmanabhan}, N.~{Palanque-Delabrouille}, K.~{Pan},
  J.~K. {Parejko}, I.~{P{\^a}ris}, W.~J. {Percival}, I.~{P{\'e}rez-Fournon},
  I.~{P{\'e}rez-R{\`a}fols}, P.~{Petitjean}, R.~{Pfaffenberger}, J.~{Pforr},
  M.~M. {Pieri}, F.~{Prada}, A.~M. {Price-Whelan}, M.~J. {Raddick},
  R.~{Rebolo}, J.~{Rich}, G.~T. {Richards}, C.~M. {Rockosi}, N.~A. {Roe}, A.~J.
  {Ross}, N.~P. {Ross}, G.~{Rossi}, J.~A. {Rubi{\~n}o-Martin}, L.~{Samushia},
  A.~G. {S{\'a}nchez}, C.~{Sayres}, S.~J. {Schmidt}, D.~P. {Schneider}, C.~G.
  {Sc{\'o}ccola}, H.-J. {Seo}, A.~{Shelden}, E.~{Sheldon}, Y.~{Shen}, Y.~{Shu},
  A.~{Slosar}, S.~A. {Smee}, S.~A. {Snedden}, F.~{Stauffer}, O.~{Steele}, M.~A.
  {Strauss}, A.~{Streblyanska}, N.~{Suzuki}, M.~E.~C. {Swanson}, T.~{Tal},
  M.~{Tanaka}, D.~{Thomas}, J.~L. {Tinker}, R.~{Tojeiro}, C.~A. {Tremonti},
  M.~{Vargas Maga{\~n}a}, L.~{Verde}, M.~{Viel}, D.~A. {Wake}, M.~{Watson},
  B.~A. {Weaver}, D.~H. {Weinberg}, B.~J. {Weiner}, A.~A. {West}, M.~{White},
  W.~M. {Wood-Vasey}, C.~{Yeche}, I.~{Zehavi}, G.-B. {Zhao}, and Z.~{Zheng},
  {\it {The Baryon Oscillation Spectroscopic Survey of SDSS-III}},  {\em \aj}
  {\bf 145} (Jan., 2013) 10, [\href{http://xxx.lanl.gov/abs/1208.0022}{{\tt
  arXiv:1208.0022}}].

\bibitem{Andersonetal:12}
L.~{Anderson}, E.~{Aubourg}, S.~{Bailey}, D.~{Bizyaev}, M.~{Blanton}, A.~S.
  {Bolton}, J.~{Brinkmann}, J.~R. {Brownstein}, A.~{Burden}, A.~J. {Cuesta},
  L.~A.~N. {da Costa}, K.~S. {Dawson}, R.~{de Putter}, D.~J. {Eisenstein},
  J.~E. {Gunn}, H.~{Guo}, J.-C. {Hamilton}, P.~{Harding}, S.~{Ho},
  K.~{Honscheid}, E.~{Kazin}, D.~{Kirkby}, J.-P. {Kneib}, A.~{Labatie},
  C.~{Loomis}, R.~H. {Lupton}, E.~{Malanushenko}, V.~{Malanushenko},
  R.~{Mandelbaum}, M.~{Manera}, C.~{Maraston}, C.~K. {McBride}, K.~T. {Mehta},
  O.~{Mena}, F.~{Montesano}, D.~{Muna}, R.~C. {Nichol}, S.~E. {Nuza}, M.~D.
  {Olmstead}, D.~{Oravetz}, N.~{Padmanabhan}, N.~{Palanque-Delabrouille},
  K.~{Pan}, J.~{Parejko}, I.~{P{\^a}ris}, W.~J. {Percival}, P.~{Petitjean},
  F.~{Prada}, B.~{Reid}, N.~A. {Roe}, A.~J. {Ross}, N.~P. {Ross},
  L.~{Samushia}, A.~G. {S{\'a}nchez}, D.~J. {Schlegel}, D.~P. {Schneider},
  C.~G. {Sc{\'o}ccola}, H.-J. {Seo}, E.~S. {Sheldon}, A.~{Simmons}, R.~A.
  {Skibba}, M.~A. {Strauss}, M.~E.~C. {Swanson}, D.~{Thomas}, J.~L. {Tinker},
  R.~{Tojeiro}, M.~V. {Maga{\~n}a}, L.~{Verde}, C.~{Wagner}, D.~A. {Wake},
  B.~A. {Weaver}, D.~H. {Weinberg}, M.~{White}, X.~{Xu}, C.~{Y{\`e}che},
  I.~{Zehavi}, and G.-B. {Zhao}, {\it {The clustering of galaxies in the
  SDSS-III Baryon Oscillation Spectroscopic Survey: baryon acoustic
  oscillations in the Data Release 9 spectroscopic galaxy sample}},  {\em
  \mnras} {\bf 427} (Dec., 2012) 3435--3467,
  [\href{http://xxx.lanl.gov/abs/1203.6594}{{\tt arXiv:1203.6594}}].

\bibitem{BOSS_project}
``\url{http://www.sdss3.org/collaboration/description.pdf}.''

\bibitem{Colomboetal:08}
L.~P.~L. {Colombo}, E.~{Pierpaoli}, and J.~R. {Pritchard}, {\it {Cosmological
  parameters after WMAP5: forecasts for Planck and future galaxy surveys}},
  {\em \mnras} {\bf 398} (Oct., 2009) 1621--1637,
  [\href{http://xxx.lanl.gov/abs/0811.2622}{{\tt arXiv:0811.2622}}].

\bibitem{Hojjati:2012ci}
A.~{Hojjati}, {\it {Degeneracies in parametrized modified gravity models}},
  {\em \jcap} {\bf 1} (Jan., 2013) 9,
  [\href{http://xxx.lanl.gov/abs/1210.3903}{{\tt arXiv:1210.3903}}].

\bibitem{Scoccimarro:04}
R.~{Scoccimarro}, {\it {Redshift-space distortions, pairwise velocities, and
  nonlinearities}},  {\em \prd} {\bf 70} (Oct., 2004) 083007,
  [\href{http://xxx.lanl.gov/abs/astro-ph/0407214}{{\tt astro-ph/0407214}}].

\bibitem{Matsubara:08}
T.~{Matsubara}, {\it {Nonlinear perturbation theory with halo bias and
  redshift-space distortions via the Lagrangian picture}},  {\em \prd} {\bf 78}
  (Oct., 2008) 083519, [\href{http://xxx.lanl.gov/abs/0807.1733}{{\tt
  arXiv:0807.1733}}].

\bibitem{Taruyaetal:10}
A.~{Taruya}, T.~{Nishimichi}, and S.~{Saito}, {\it {Baryon acoustic
  oscillations in 2D: Modeling redshift-space power spectrum from perturbation
  theory}},  {\em \prd} {\bf 82} (Sept., 2010) 063522,
  [\href{http://xxx.lanl.gov/abs/1006.0699}{{\tt arXiv:1006.0699}}].

\bibitem{HikageYamamoto:13}
C.~{Hikage} and K.~{Yamamoto}, {\it {Impacts of satellite galaxies in measuring
  the redshift-space distortions}},  {\em ArXiv e-prints} (Mar., 2013)
  [\href{http://xxx.lanl.gov/abs/1303.3380}{{\tt arXiv:1303.3380}}].

\bibitem{Hikageetal:12a}
C.~{Hikage}, M.~{Takada}, and D.~N. {Spergel}, {\it {Using galaxy-galaxy weak
  lensing measurements to correct the finger of God}},  {\em \mnras} {\bf 419}
  (Feb., 2012) 3457--3481, [\href{http://xxx.lanl.gov/abs/1106.1640}{{\tt
  arXiv:1106.1640}}].

\end{thebibliography}\endgroup

\end{document}